\documentclass[11pt,a4paper]{article}
\usepackage{jheppub}
\usepackage{mathtools}
 \usepackage{comment}

\usepackage{color}

\newcommand{\be}{\begin{equation}}
\newcommand{\ee}{\end{equation}}
\newcommand{\bea}{\begin{eqnarray}}
\newcommand{\eea}{\end{eqnarray}}
\newcommand{\nn}{\nonumber}

\newcommand{\9}[1]{\textsl{#1}}

\newcommand{\n}{_{n=1}^\infty}
\newcommand{\dz}{_0^{z_0}dz}

\newcommand{\C}{\mathcal{C}}
\newcommand{\F}{\mathcal{F}}
\newcommand{\J}{\mathcal{J}}
\newcommand{\cM}{\mathcal{M}}
\newcommand{\beq}{\begin{equation}}
\newcommand{\eeq}{\end{equation}}
\newcommand{\cF}{\mathcal{F}}
\newcommand{\Q}{\mathcal{Q}}
\newcommand{\M}{\mathcal{M}}

\newcommand{\QQ}{\text{tr}\Q^2}

\def\CGN{\cite{Colangelo:2024xfh}}
\def\HS{\cite{Hoferichter:2020lap}}

\def\tHSZ{\cite{Hoferichter:2024bae}}

\newcommand{\toprule}{\hline}
\newcommand{\colrule}{\hline}
\newcommand{\botrule}{\hline}

\begin{document}

\title{Tensor meson transition form factors\\ in holographic QCD
and the muon $g-2$
}

\author[a,b]{Luigi Cappiello,}
\affiliation[a]{Dipartimento di Fisica ``Ettore Pancini", Universit\`a di Napoli ``Federico II"} 
\affiliation[b]{
INFN-Sezione di Napoli, Via Cintia, I-80126 Napoli, Italy}
\author[c]{Josef Leutgeb,}
\author[c]{Jonas Mager}
\author[c]{and Anton Rebhan}
\affiliation[c]{Institut f\"ur Theoretische Physik, TU Wien,
        Wiedner Hauptstrasse 8-10, A-1040 Vienna, Austria}


\abstract{%
Despite the prominence of tensor mesons in photon-photon collisions, until
recently their contribution to the hadronic light-by-light 
(HLbL)
scattering part of
the anomalous magnetic moment of the muon
has been estimated to be at the level of only a few $10^{-12}$,
with an almost negligible contribution to the error budget of the Standard Model prediction. A recent reanalysis within the dispersive approach has found that
after resolving the issue of kinematic singularities in previous approaches,
a larger result
is obtained, a few $10^{-11}$, and with opposite sign as in previous results,
when a simple quark model for the transition form factors is employed.
In this paper, we present the first complete evaluation of tensor meson
contributions within a hard-wall model in holographic QCD, which
reproduces surprisingly well mass, two-photon width, and 
the observed singly virtual transition form factors
of the dominant $f_2(1270)$, requiring only that
the energy-momentum tensor correlator is matched to the leading OPE result of QCD.
Due to a second structure function that is absent in the quark model
and in lowest-order resonance chiral theory,
the result for $a_\mu$ turns out to be positive instead of negative, 
and also with a magnitude of a few $10^{-11}$. 
We discuss both
pole and non-pole contributions arising from tensor meson exchanges
in the holographic HLbL amplitude, finding
that keeping all contributions improves dramatically the convergence
of a sum over excited tensor mesons and avoids unnaturally large
contributions from the first few excited modes at low energies.
Moreover, we find that the infinite tower of tensor mesons permits to fill the
gap in the symmetric longitudinal short-distance constraint on the HLbL amplitude
left by the contribution of axial vector mesons.
Matching the corresponding leading-order OPE result leads to two-photon couplings consistent with the observed
combined effects
of the ground-state $f_2,a_2,f_2'$ multiplet and a total $a_\mu^\mathrm{Tensor}$ contribution
of $+12.4\times 10^{-11}$; with an $F_\rho$ fit this is reduced slightly to $+11.1\times 10^{-11}$.
A contribution of this size from the tensor sector could
explain the tension between the most recent dispersive and
lattice results for $a_\mu^\mathrm{HLbL}$.
}

\maketitle

\section{Introduction}

With the upcoming release of the final result of the Fermilab experiment
measuring the anomalous magnetic moment of the muon that is expected to reduce the current experimental error of $22\times 10^{-11}$ in $a_\mu=(g-2)_\mu/2$ by about a factor of 2, there is currently a world-wide effort
to reduce also the uncertainties of the Standard Model prediction which are
dominated by hadronic contributions, foremost from hadronic vacuum polarization (HVP),
but also from hadronic light-by-light scattering (HLbL) with an error budget of $19\times 10^{-11}$
according to the 2020 White Paper of the Muon $g-2$ Theory Initiative \cite{Aoyama:2020ynm}.

In the meantime, significant progress has been made regarding
the various parts of the HLbL amplitude, in particular regarding the
contribution of axial vector mesons 
and short distance constraints \cite{Melnikov:2003xd,Colangelo:2019lpu,Colangelo:2019uex,Leutgeb:2019gbz,Cappiello:2019hwh,Knecht:2020xyr,Masjuan:2020jsf,Ludtke:2020moa,Bijnens:2020xnl,Bijnens:2021jqo,Colangelo:2021nkr,Bijnens:2022itw,Bijnens:2024jgh}.
Different approaches, involving dispersion relations \cite{Colangelo:2014dfa,Colangelo:2014pva,Colangelo:2015ama,Colangelo:2017fiz}, Dyson-Schwinger/Bethe-Salpeter
equations \cite{Eichmann:2019tjk, Eichmann:2019bqf, Eichmann:2024glq}, resonance chiral models \cite{Roig:2014uja,Guevara:2018rhj,Estrada:2024cfy}, as well as holographic QCD  \cite{Leutgeb:2019gbz,Cappiello:2019hwh,Leutgeb:2021mpu,Leutgeb:2022lqw, Leutgeb:2024rfs} have
been employed with results that are sufficiently in agreement to
permit an improved estimate with substantially reduced theoretical errors.

A contribution where a rigorous dispersive analysis is not yet available
is the one of tensor mesons, where $f_2(1270)$ and $a_2(1320)$ have
sufficiently strong coupling to two photons such that they should be
taken into account, since they are too light to be accounted for
by the quark-loop contribution which provides sufficiently good accuracy for
virtualities higher than about 1.5 GeV.

In \cite{Danilkin:2016hnh} the contribution of the tensor mesons $f_2(1270)$ and $a_2(1320)$ has been estimated with a quark model ansatz for the transition form factor
(TFF) \cite{Schuler:1997yw}
as $+0.50(13)\times 10^{-11}$ and $+0.14(3)\times 10^{-11}$, respectively,
and a very similar result of $+0.6\times 10^{-11}$ has recently been
obtained by using holographic QCD as a model in \CGN, however in
an expressly incomplete evaluation.
Recently, larger results with an opposite sign have been obtained
in a new dispersive analysis using a formalism that avoids
the kinematic singularities present in previous approaches 
\cite{Hoferichter:2024bae}, amounting to
$-2.5(3)\times 10^{-11}$ in the low-energy region bounded by 1.5 GeV.

In this paper, we present our results for a complete evaluation of the
HLbL contribution obtained by employing the results for
the TFF of tensor mesons in a 
holographic QCD (hQCD)
hard-wall (HW) model, which has been
found to work well in other applications involving TFFs of pseudoscalars and
axial vector mesons. As  already found in \cite{Katz:2005ir},
a hard-wall AdS/QCD model where the energy-momentum tensor 
correlator is matched to the leading OPE result of QCD
reproduces the mass of the dominant $f_2(1270)$ within 3\%,
and its two-photon width completely within experimental errors.
As already shown in \CGN\ for the helicity-2 amplitude, and extended here
to all helicities,
the singly virtual transition form factors
agree quite well with data from the Belle collaboration \cite{Belle:2015oin}.
Using  the complete set of tensor TFFs
 obtained in hQCD in the formulae of the dispersive approach,
we obtain a significantly larger positive 
result than the holographic study of ref.~\CGN,
larger in absolute value also than the result of ref.~\cite{Hoferichter:2024bae}.

We also consider a full evaluation beyond the pole contribution
as defined by the dispersive approach with optimized basis \cite{Hoferichter:2024fsj},
by keeping the complete HLbL amplitude as given by the holographic model.
In contrast to the case of pseudoscalar and axial vector mesons, where
the complete HLbL amplitude yields an $a_\mu$ contribution identical to
the dispersively defined pole contribution, for tensor mesons
the resulting formulae differ, and an even larger positive result is obtained.

The holographic model also provides  an infinite tower of tensor meson
resonances, and we find that summing over these contributions
gives a still somewhat larger result, with the bulk of the contribution still
coming from the region below 1.5 GeV. Summing over the first few modes, we
find a similar total result, but if this sum is carried out
with only the pole contribution, the convergence is rather slow,
with unnaturally large contributions from the excited tensor modes
even at low energies.
Moreover, we find that the infinite sum of tensor modes
contributes to the symmetric longitudinal short-distance limit
of the HLbL amplitude, but not to the asymmetric one involved
in the Melnikov-Vainshtein constraint \cite{Melnikov:2003xd}.
In holographic models it was shown that the latter is saturated by the infinite tower of axial vector mesons \cite{Leutgeb:2019gbz,Cappiello:2019hwh,Leutgeb:2021mpu,Leutgeb:2021bpo,Leutgeb:2022lqw} while the symmetric longitudinal short-distance
constraint \cite{Colangelo:2019lpu,Colangelo:2019uex,Bijnens:2020xnl,Bijnens:2021jqo}
is matched only at the level of 81\%.
With a normalization of the tensor modes such that this gap is filled
by tensor mesons, one can correct the shortcoming of the original
prescription of \cite{Katz:2005ir} that at large $N_c$ the $f_2$ coupling scales
like that of a tensor glueball instead of a quarkonium state, and
numerically one obtains a two-photon coupling of the lightest tensor
mode that agrees well with the observed coupling of the full
ground-state multiplet $f_2,a_2,f_2'$.

This paper is organized as follows. In sec.~\ref{sec:generalTFFs}, we
recall the basic formulae for transition form factors of tensor mesons,
following the notation of \HS. In sec.~\ref{sec:AdSQCD}, we set up
the hard-wall AdS/QCD model used by us and we fix its parameters, which
involve only the size of the extra dimension related to the value of the $\rho$ mass, a five-dimensional flavor-gauge theory
coupling fixed by matching to OPE of the vector-vector correlation function,
and a five-dimensional Newton constant fixed by matching
the OPE of the energy-momentum-tensor correlator in $N_f=3$ QCD.
In sec.~\ref{sec:hTFF} we derive the holographic result for the tensor meson
TFFs and  compare with the experimental data of \cite{Belle:2015oin},
before evaluating in sec.~\ref{sec:amuT}
both the $a_\mu^\mathrm{HLbL}$ contribution resulting from the
singlet tensor meson as it is naturally present in the model and also
when used as a model with refitted masses and two-photon widths to
match $f_2(1270)$, $a_2(1320)$, and $f_2'(1525)$.
The $a_\mu$ contributions are evaluated both with a restriction to
the pole term as defined by the dispersive approach in the optimized
basis of ref.~\cite{Hoferichter:2024fsj} and without.
With the latter we find a quicker convergence to the full sum
over excited tensor mesons, avoiding unnaturally large contributions
from the first few excited states.
The infinite sum of tensor contributions is then evaluated by means of
the tensor bulk-to-bulk propagator, with which we obtain
a nonvanishing contribution to the symmetric longitudinal short-distance
limit of the HLbL amplitude. From this we fix the normalization of the
tensor modes and evaluate the contribution to $a_\mu$ for two
alternative schemes, a fit to the leading-order OPE result and
one where the asymptotic limit is reached only to 89.4\%, corresponding
to a reduced
five-dimensional gauge
coupling that fits the $\rho$ decay constant $F_\rho$
instead of the OPE result for the vector correlator.
The resulting contributions to $a_\mu$ are evaluated
with a breakup into low, high, and mixed energy regions.
Appendix \ref{sec:modedecomposition} discusses the decomposition of the
holographic TFFs in vector meson modes and their relation to resonance chiral theory \cite{Ecker:1988te,Ecker:1989yg},
as well as how complete vector meson dominance (VMD) is realized in hQCD.
Full expressions for the HLbL amplitude components are given
for single modes and the complete tower in Appendices \ref{appB}
and \ref{appC}, respectively.

\section{Transition form factors of tensor mesons}\label{sec:generalTFFs}

The matrix element of a massive tensor meson decaying into two off-shell photons 
is given by
\begin{align}
&\langle \gamma^*(q_1,\lambda_1)\gamma^*(q_2,\lambda_2) | T(p,\lambda_T) \rangle =\\
&i(2\pi)^4 \delta^{(4)}(q_1+q_2-p) e^2 {\epsilon_\mu^{\lambda_1}}^*(q_1) {\epsilon_\nu^{\lambda_2}}^*(q_2)\epsilon_{\alpha\beta}^{\lambda_T}(p) \cM^{\mu\nu\alpha\beta}(q_1,q_2),&\nonumber
\end{align} 
with
\begin{align}
&\epsilon_{\alpha\beta}^{\lambda_T}(p) \cM^{\mu\nu\alpha\beta}(q_1,q_2) \nonumber\\
&= i\int d^4x \, e^{i q_1 \cdot x} \langle 0 | T \{ j_\mathrm{em}^\mu(x) j_\mathrm{em}^\nu(0) \} | T(p,\lambda_T) \rangle .
\end{align}
The expressions of  the massive tensor polarization  $\epsilon_{\alpha\beta}^{\lambda_T}$ can be found in \cite{Poppe:1986dq,Hoferichter:2020lap}. The  sum over polarizations gives the projector
\begin{align}
	\label{TensorPolarizationSum}
	s^T_{\alpha\beta\alpha'\beta'}(p) &= \sum_{\lambda_T} \epsilon_{\alpha\beta}^{\lambda_T}(p) \epsilon_{\alpha'\beta'}^{\lambda_T}(p)^* \nonumber\\
	&= \frac{1}{2} \left( s_{\alpha\beta'} s_{\alpha'\beta} + s_{\alpha\alpha'} s_{\beta\beta'} \right) - \frac{1}{3} s_{\alpha\beta} s_{\alpha'\beta'} ,
\end{align}
where
\beq
	s_{\alpha\alpha'} = - \left( g_{\alpha\alpha'} - \frac{p_\alpha p_{\alpha'}}{m_T^2} \right),
\eeq
satisfies
\beq
	\eta^{\alpha'\alpha''} \eta^{\beta'\beta''} s^T_{\alpha\beta\alpha'\beta'} s^T_{\alpha''\beta''\alpha'''\beta'''}  =  s^T_{\alpha\beta\alpha'''\beta'''}, 
\eeq
and enters in the expression of the massive tensor (Fierz-Pauli) propagator
\beq
\label{eq:FP-prop}
G^T_{\alpha\beta,\alpha'\beta'}=\frac{s^T_{\alpha\beta\alpha'\beta'}(p)}{p^2-m_T^2}.
\eeq
In the literature, there are  two widely used  choices of basis of tensor structures. One is aimed at explicitly selecting amplitudes of given helicity \cite{Poppe:1986dq}, more directly related to experiments, while the second choice follows the so called BTT construction \cite{Bardeen:1969aw,Tarrach:1975tu}, used in the data driven dispersive approach,  which aims at obtaining a decomposition of the transition form factors, with  scalar coefficients  not having kinematical singularities \cite{Hoferichter:2020lap}. 

Following the second approach, Lorentz and gauge invariance with respect to photon momenta, crossing symmetry requirement
\beq
	\cM^{\mu\nu\alpha\beta}(q_1,q_2) = \cM^{\nu\mu\alpha\beta}(q_2,q_1) , 
\eeq
and the observation that  only those structures that do not vanish upon contraction with the projector $s^T_{\alpha\beta\alpha'\beta'}$ can contribute to observables
involving on-shell tensor mesons narrows the choice to five independent tensor structures (Levi-Civita  tensor structures are excluded by parity conservation):
\bea
T_1^{\mu\nu\alpha\beta} &=& g^{\mu\alpha} P_{21}^{\nu\beta} + g^{\nu\alpha} P_{12}^{\mu\beta} + g^{\mu\beta} P_{21}^{\nu\alpha}\nonumber\\
& +& g^{\nu\beta} P_{12}^{\mu\alpha} + g^{\mu\nu} (q_1^\alpha q_2^\beta + q_2^\alpha q_1^\beta) \nonumber\\
& -& q_1 \cdot q_2 ( g^{\mu\alpha} g^{\nu\beta} + g^{\nu\alpha} g^{\mu\beta} )  ,\nonumber \\
		T_2^{\mu\nu\alpha\beta} &=& (q_1^\alpha q_1^\beta + q_2^\alpha q_2^\beta ) P_{12}^{\mu\nu} , \nonumber\\
		T_3^{\mu\nu\alpha\beta} &=& P_{11}^{\mu\alpha} P_{22}^{\nu\beta} + P_{11}^{\mu\beta} P_{22}^{\nu\alpha} , \\
		T_4^{\mu\nu\alpha\beta} &=& P_{12}^{\mu\alpha} P_{22}^{\nu\beta} + P_{12}^{\mu\beta} P_{22}^{\nu\alpha} , \nonumber\\
		T_5^{\mu\nu\alpha\beta} &= &P_{21}^{\nu\alpha} P_{11}^{\mu\beta} + P_{21}^{\nu\beta} P_{11}^{\mu\alpha} ,\nonumber
	\label{structures_tensors}
\eea
where
\beq
	P_{ij}^{\mu\nu} = g^{\mu\nu} q_i \cdot q_j - q_i^\nu q_j^\mu .
\eeq
Thus, the tensor TFF depends on five scalar coefficients 
$\cF_i(q_1^2,q_2^2)$:
\beq\label{5TensorTFF}
	\cM^{\mu\nu\alpha\beta} = \sum_{i=1}^5  \frac{\cF_i^T}{m_T^{n_i}} T_i^{\mu\nu\alpha\beta},
\eeq
where $n_1=1, n_i=3\; \text{for}\; j=2,...,5$, so that all the $\cF_i$ are dimensionless.

Only $\cF_1$ and $\cF_2$ enter the  on-shell photon result
\beq
\label{GammaTgammagamma}
	\Gamma_{T\rightarrow\gamma\gamma} = \frac{\pi \alpha^2 m_T}{5} \left( | \cF_1^T(0,0) |^2  + \frac{1}{24m_T^6} | \cF_2^T(0,0) |^2 \right) .
\eeq
In singly virtual TFFs, all $\F^T_i$ contribute, except for
$\F^T_3$, unless the latter has a singularity at zero virtualities.

In previous studies of the contribution of tensor mesons
to HLbL scattering and to $a_\mu$, only the simple
quark model ansatz of ref.~\cite{Schuler:1997yw} for $\F^T_1$,
with the remaining $\F^T_i$ set to zero, has been employed:
\be\label{eq:FT1QM}
\F^T_1(-Q_1^2,-Q_2^2)=\frac{\F^T_1(0,0)}{(1+(Q_1^2+Q_2^2)/\Lambda_T^2)^2}.
\ee
In the following we shall employ holographic QCD as a model
for the TFFs of mesons and compare with the results obtained
by the quark model with common choices of $\Lambda_T$.

\section{AdS/QCD}\label{sec:AdSQCD}

Holographic QCD models \cite{Sakai:2004cn,Erlich:2005qh,DaRold:2005mxj, Hirn:2005nr, Karch:2006pv} have been constructed along the lines of the original conjectured AdS/CFT duality (equivalence)  between a  four-dimensional (4D) (conformal) large-N$_c$ gauge theory at strong coupling and a (classical) five-dimensional  (5D) field theory in a curved gravitational background with Anti-de-Sitter metric \cite{Maldacena:1997re}, which can be summarized as follows~\cite{Gubser:1998bc,Witten:1998qj}: for every quantum
operator ${\cal O}(x)$ of the 4D (strongly coupled) gauge theory, there exists a corresponding 5D 
field $\phi(x,z)$, whose value on the conformal boundary (taken at $z=\epsilon\to0$)
$\phi(x,0)\equiv\phi_0(x)$, is identified, modulo some specific powers of $\epsilon$, with the
four-dimensional source of ${\cal O}(x)$. The generating
functional of the 4D theory can be computed from the 5D action
evaluated \emph{on-shell}, {\it{i.e.}}:
\begin{equation}
\mbox{exp}\left(i
S_5[\phi_0(x)]\right)=\left\langle\mbox{exp}\left[i\int d^4 x
\,\phi_0(x)\,{\cal O}(x)\right]\right\rangle_{\rm
QCD_4}~.\label{holography}
\end{equation}
By varying the action with respect to the 4D boundary values $\phi_0(x)$, one  generates
 connected $n$-point Green's functions of large-$N_c$, strong-coupled 4D gauge theory.

We shall introduce a 5D tensor field as a metric deformation in  hQCD models together with 5D gauge fields which are used to compute 
correlators of (conserved) $SU(3)_L\times SU(3)_R$ (or
$U(3)_L\times U(3)_R$) flavor chiral currents of QCD, in the large-N$c$ limit,
\beq J^a_{L\,\mu}= \overline{q}_L \, \gamma^{\mu}\,T^a \,
q_L,\quad J^a_{R\,\mu}= \overline{q}_R \, \gamma^{\mu}\,T^a \, q_R
\label{chiralcurrents}\eeq
where $T^a=\lambda^a/2$, with $\lambda^a$, $a=1,...8$ being the
$SU(3)$ Gell-Mann matrices, augmented by $T^0=\textbf{1}_3/\sqrt{6}$, such that
$\text{tr}\,(T^a \,T^b)=\delta_{ab}/2$. 

The 5D action describes a Yang-Mills theory (with a Chern-Simons term which we omit, because it plays no role here) in a curved 5D AdS$_5$ space, with the extra dimension $z$, extending over the finite interval $(0,z_0]$, and metric
\be
ds^2=z^{-2}(\eta_{\mu\nu}dx^\mu dx^\nu - dz^2).
\ee
The 5D Yang-Mills action is given by 
\be\label{eq:SYM5D}
S_{\rm YM} = -\frac{1}{4g_5^2} \int d^4x \int_0^{z_0} dz\,
\sqrt{-g}\, g^{PR}g^{QS}
\,\text{tr}\left(\mathcal{F}^\mathrm{L}_{PQ}\mathcal{F}^\mathrm{L}_{RS}
+\mathcal{F}^\mathrm{R}_{PQ}\mathcal{F}^\mathrm{R}_{RS}\right),
\ee
where $P,Q,R,S=0,\dots,3,z$ and $\mathcal{F}_{MN}=\partial_M \mathcal{B}_N-\partial_N \mathcal{B}_M-i[\mathcal{B}_M,\mathcal{B}_N]$, $\mathcal{B}_N=L_N,\;R_N$ being 5D gauge
fields  transforming under $U(3)_{L, R}$ respectively. Vector and axial-vector fields are given by ${V}_{\mu}=\frac{1}{2}( {L}_{\mu} + {R}_{\mu})$ and $ {A}_{\mu}=\frac{1}{2}( {L}_{\mu} - {R}_{\mu})$. 
Boundary conditions have to be imposed on the 5D vector fields at 
the hard wall at
$z=z_0$
which breaks conformal invariance in the infrared and implements
confinement. 

For the vector fields, working in the $V_z=0$ gauge and imposing the boundary conditions $V_\mu(x,0)=0$ and $\partial_z V_\mu(x,z_0)=0$,  the expansion $V_\mu^a(x,z)=g_5\sum_{n=1}^\infty v_\mu^{(n)a}(x)\psi_n(z)$ is obtained in terms of 4D canonically normalized vector fields $v_\mu^a(x)$, where the $\psi_n(z)$ are solutions of the 5D equation of motions, vanishing at $z=0$ and such that $\partial_z\psi_n(z)=0$, with normalization $\int_0^{z_0} dz\,z^{-1}\psi_n(z)^2=1$, explicitly:
\be\label{eq:psin}
\psi_n(z)=\sqrt{2}\frac{z J_1(\gamma_{0,n}z/z_0)}{z_0J_1(\gamma_{0,n})}.
\ee
 Masses are given in terms of the zeros of the Bessel function $J_0(\gamma_{0,n})=0$  by $m_n=\gamma_{0,n}/z_0$.

Identifying the mass of the lowest vector resonance with the mass of the $\rho$ meson $m_1=\gamma_{0,1}/z_0=M_\rho=775.26$ MeV fixes
\be\label{eq:z0}
z_0=\gamma_{0,1}/m_1=3.102 \, \text{GeV}^{-1}.
\ee
A fundamental object in a holographic model is the bulk-to-boundary propagator, which is a solution of the vector 5D equation of motion, with $m_n^2$ replaced by $q^2$ and the boundary conditions 
$\J(Q,0)=1$ and $\partial_z \J(Q,z_0)=0$. 
For Euclidean momenta $q^2=-Q^2$, it is given by 
\be\label{HWVF}
\J(Q,z)=
Qz \left[ K_1(Qz)+\frac{K_0(Q z_0)}{I_0(Q z_0)}I_1(Q z) \right],
\ee
where $I$ and $K$ denote modified Bessel functions.

Using the holographic recipe, the 4D  vector current two-point function can be written in terms of (\ref{HWVF}), and matching 
the leading-order pQCD result \cite{Erlich:2005qh},
\be\label{PiVas}
\Pi_V(Q^2)=-\frac1{g_5^2 Q^2} \left( \frac1z \partial_z \J(Q,z) \right)\Big|_{z\to0}=-\frac{N_c}{24\pi^2}\ln Q^2 + \ldots,
\ee
which determines $g_5^2=12\pi^2/N_c$. 

As already mentioned above, following \cite{Katz:2005ir}, the tensor meson is introduced in the model as a deformation of the 4D part of the $AdS$ metric
\begin{equation}\label{eq:metricfluctuation}
    ds^2=g_{MN}dx^M dx^N =\frac{1}{z^2} (\eta_{\mu\nu}+h_{\mu\nu}) dx^\mu dx^\nu-\frac{1}{z^2} dz^2\,.
\end{equation}
Its dynamics is described by the 5D Einstein-Hilbert action. 
\bea
    S_\mathrm{EH} &=& -2k_T\int d^5x \,\sqrt{g} \, (\mathcal{R}+2\Lambda)\nonumber\\
    &=& -\frac{k_T}{2} \int d^5x \, \frac{1}{z^3} \eta^{\mu\alpha} \eta^{\nu\beta} (\partial_z h_{\mu\nu} \partial_z h_{\alpha\beta} + h_{\mu\nu} \Box h_{\alpha\beta}+\ldots)\,,
\eea

It is possible to  fix the value of constant $k_T$ in a manner analogous to what was  done for $g_5$, making the assumption that $h_{\mu\nu}(x,z)$ couples on the conformal boundary $z=0$ to the energy-momentum tensor of QCD fields.  Then, the two-point function of the 4D energy-momentum tensor can be identified as\footnote{Ref.~\CGN\ is missing the factor 4 arising from $T^{\mu\nu}=2\times\delta S/\delta h_{\mu\nu}$.\label{fn:factor4}}
\be
\langle T^{\mu\nu}T^{\rho\sigma} \rangle
=4 \frac{\delta^2 S_\mathrm{on-shell}}{\delta h_{\mu\nu}\delta h_{\rho\sigma}}
\to -\frac12 k_T P^{\mu\nu\rho\sigma} Q^4 \ln Q^2
\ee
with the transverse-traceless projector
$P_{\mu\nu\rho\sigma}=\frac{1}{2}(P_{\mu\rho}P_{\nu\sigma}+P_{\mu\sigma}P_{\nu\rho}-\frac{2}{3}P_{\mu\nu}P_{\rho\sigma})$
with $P_{\mu\nu}=\eta_{\mu\nu}-q_\mu q_\nu/q^2$.
Matching to the
leading term of the OPE of the energy-momentum tensor correlator \cite{Novikov:1981xi}
\be
\langle T^{\mu\nu}T^{\rho\sigma} \rangle_\mathrm{QCD}
=-P^{\mu\nu\rho\sigma} \left(\frac{N_c N_f}{160\pi^2}+\frac{N_c^2-1}{80\pi^2}\right) Q^4 \ln Q^2+\ldots
\ee
determines \cite{Katz:2005ir}
\be\label{kTpQCD}
k_T=\left(\frac{N_c N_f}{80\pi^2}+\frac{N_c^2-1}{40\pi^2}\right).
\ee

Solutions of traceless-transverse metric fluctuation 
$h_{\mu\nu}=\epsilon_{\mu\nu}^T(q)h_n(z)$
with $h_n(0)=0=h_n'(z_0)$, satisfying the canonical normalization ${k_T} \int dz z^{-3} (\partial_z h_n)^2={(m^T_n)^2}$, are given by
\beq
    h_n(z) = \displaystyle\frac{\sqrt{2/k_T}}{{z_0}
   J_2\left(\gamma_{1,n}\right)} z^2 J_2(\gamma_{1,n}z/z_0),
\eeq
yielding a tower of (singlet) tensor meson modes
with masses fixed by
\be
m_n^T/M_\rho=\gamma_{1,n}/\gamma_{0,1}=
1.593,\, 2.917,\, 4.230,\, \ldots 
\ee
The lowest tensor resonance mass is thus $m^T_1=1.235$ GeV, differing  by only about $3\%$ from the physical value $M_{f_2}=1.2754$ GeV of $f_2(1270)$.\footnote{To
distinguish experimental values of meson masses from model parameters, we
use upper-case $M$ for the former.}

\section{Holographic tensor meson TFF }\label{sec:hTFF}

Photons with polarization vector $\epsilon^\mu(q)$, momentum $q$ and virtuality $Q^2=-q^2$
are described by a vector flavor gauge field involving the
vector bulk-to-boundary propagator $\J(Q,z)$ according to (for $N_f=3$)
\be
V^\mu=\Q\,\J(Q,z)\epsilon^\mu(q),\quad
\Q=e\;\mathrm{diag}\left(\frac23,-\frac13,-\frac13\right).
\ee

Expanding the action \eqref{eq:SYM5D} with deformed metric \eqref{eq:metricfluctuation} to linear order in $h_{\mu\nu}$ leads to the 5D interaction vertex given by\footnote{We differ here by a relative sign between the two tensor structures from \CGN.\label{fn:sign}}
\begin{align}\label{eq:SYMhvv}
    S_\mathrm{YM}=-\frac{1}{2 g_5^2} \text{tr} \Q^2 \int \frac{d^4q_1}{(2 \pi)^4} \frac{d^4q_2}{(2 \pi)^4}\int_0^{z_0} \frac{dz}{z} h^{\alpha\beta}(z,-q_1-q_2) \epsilon_{\mu_1 }(q_1)\epsilon_{\mu_2 }(q_2) \nonumber \\ \times\left( (T_3)^{\mu_1 \mu_2}_{\;\;\;\;\;\;\;\; \alpha\beta} \frac{\partial_z \J(Q_1,z)}{q_1^2}\frac{\partial_z \J(Q_2,z)}{q_2^2} +(T_1)^{\mu_1 \mu_2}{}_{\alpha\beta} \J(Q_1,z)\J(Q_2,z)\right).
\end{align}
For a given single tensor mode this yields the $T\to\gamma^*\gamma^*$
amplitude
\begin{align}
    \M^{\mu \nu \alpha \beta}= T_1^{\mu \nu \alpha \beta} \frac{1}{m_T}\mathcal{F}_1+ T_3^{\mu \nu \alpha \beta}\frac{1}{m_T^3} \mathcal{F}_3
\end{align}
with 
\begin{align}
    \mathcal{F}^T_1(-Q_1^2,-Q_2^2)/m_T&=- \frac{1}{g_5^2} \text{tr}\Q^2\int_0^{z_0} \frac{dz}{z} h_n(z) \J(Q_1,z)\J(Q_2,z), \label{F1}\\
       \mathcal{F}^T_3(-Q_1^2,-Q_2^2)/m_T^3&=-\frac{1}{g_5^2} \text{tr}\Q^2\int_0^{z_0} \frac{dz}{z} h_n(z) \frac{\partial_z\J(Q_1,z)}{Q_1^2} \frac{\partial_z\J(Q_2,z)}{Q_2^2},\label{F3}
\end{align}
and $\J(Q,z)$ given in \eqref{HWVF}.
Note that for real photons $\J(0,z)\equiv 1$ and that
$\partial_z\J(Q,z)\sim Q^2$ in the limit $Q^2\to0$, rendering $\F^T_3(0,0)$ finite.

The above structure is in fact universal for holographic models when the
tensor meson is described by a metric fluctuation.
In a soft-wall model \cite{Mamedov:2023sns,Colangelo:2024xfh} 
expressions analogous to \eqref{F1} and \eqref{F3} would be obtained, the only differences being that the integrals on $z$ would extend to infinity and contain   a dilaton factor and  different  $h_n(z)$ and $\J(Q,z)$.
The simple quark model TFF \eqref{eq:FT1QM} posited in \cite{Schuler:1997yw} has only $\F^T_1$.
The structure function $\F^T_3$ arises in so-called minimal models
of tensor mesons \cite{Mathieu:2020zpm} from gauge-nonvariant $h_{\mu\nu}V^\mu V^\nu$
couplings, corresponding to a nonvanishing $\F^T_3$ upon projection
on transverse modes, which would however be singular at vanishing
virtualities. In such a model, $\F^T_3$ actually contributes to the real-photon rate.
The holographic model however involves $\partial_z\J(Q,z)$ which vanishes for
real photons such that in the 4D Lagrangian the term $h_{\mu\nu}V^\mu V^\nu$ is
present only for massive vector bosons and virtual photons. This is in fact
analogous to how the Landau-Yang theorem \cite{Landau:1948kw,Yang:1950rg}
for axial vector mesons is realized
in hQCD. There the TFF involves $\J_1\partial_z \J_2$ so that at least one
photon has to be off-shell. In the case of $\F^T_3$, two factors of $\partial_z \J$ appear,
requiring double virtuality for it to contribute in decay amplitudes.\footnote{%
In appendix \ref{sec:modedecomposition} we discuss the mode decomposition of the holographic
TFFs and we show that using a representation of vector mesons by antisymmetric tensor fields
it is in fact straightforward to achieve a regular $\F_3^T$ at vanishing virtualities and thus
intact gauge invariance also with a finite number of vector meson modes.
}

Evaluating at $\mathcal{F}^T_1(0,0)$ for the lightest tensor mode 
and  using \eqref{GammaTgammagamma}, we exactly reproduce the result of 2.7 keV for $\Gamma_{T\to\gamma\gamma}$ obtained in \cite{Katz:2005ir}, when  $m_T$ is raised to the experimental value of the physical mass  $1275.5$ MeV of $f_2(1270)$; otherwise, using the mass $1235$ MeV, as predicted by the model, one obtains $2.46$ keV. This agrees with \CGN, if one corrects for the factor of 4 mentioned
in footnote \ref{fn:factor4}; the different sign in \eqref{eq:SYMhvv} does not matter for real photons.

However, this good agreement depends on the choice of $k_T$ being fixed by
\eqref{kTpQCD}, which involves the full energy-momentum tensor. In the large-$N_c$
limit this implies that the coupling constants of  tensor modes, which
are proportional to $1/\sqrt{k_T}$, go like $1/N_c$, which is characteristic
of a glueball mode, and not of a quarkonium state, which should go like $1/\sqrt{N_c}$.
In the following, we shall first consider the original choice of ref.~\cite{Katz:2005ir},
eq.~\eqref{kTpQCD},
which leads to this remarkable agreement of the two-photon width of $f_2(1270)$,
and will reconsider this choice after having inspected the short-distance
behavior of the HLbL amplitude.

\begin{figure}[h]
\centering
\includegraphics[width=0.675\textwidth]{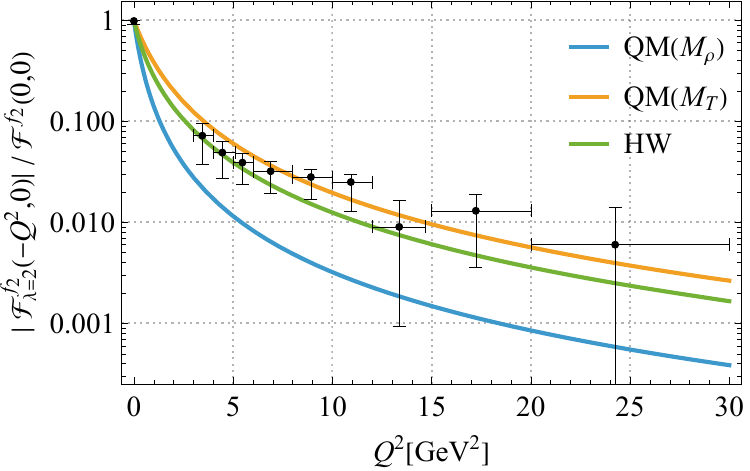}
\includegraphics[width=0.675\textwidth]{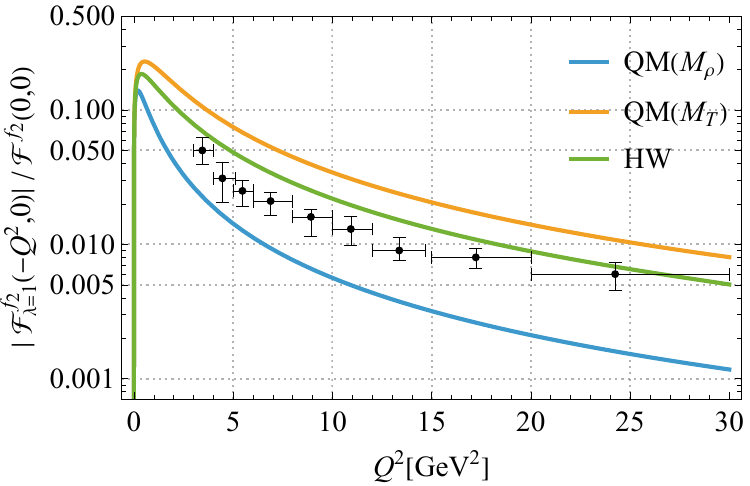}
\includegraphics[width=0.675\textwidth]{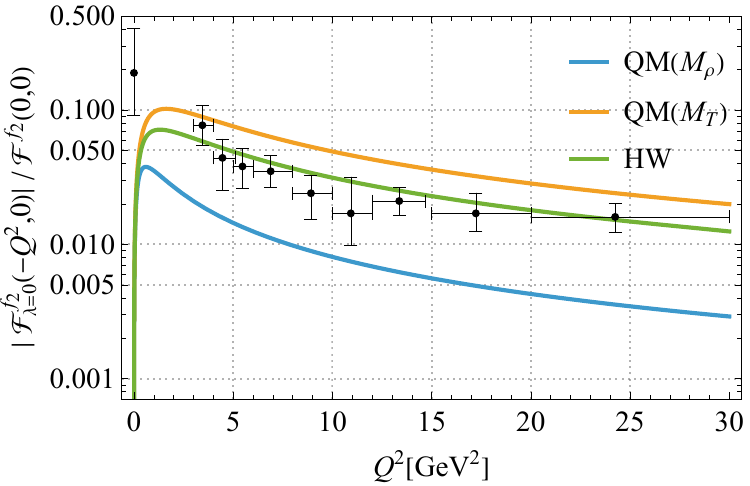}
\caption{Comparison of singly virtual tensor TFFs for helicities $\lambda=2,1,0$ with Belle data \cite{Belle:2015oin} for the $f_2(1270)$, normalized by
$\F^{f_2}(0,0)\equiv\sqrt{\F^T_1(0,0)^2+\F^T_2(0,0)^2/24}=\sqrt{5 \Gamma_{\gamma\gamma}/(\pi\alpha^2 M_T)}$.
\label{fig:belle}}
\end{figure}

\subsection{Comparison with TFF data}

The above results for the tensor TFFs have a simple overall dependence on
the parameters $g_5$ and $k_T$, which drop out if normalized TFFs are
considered. The latter only depend on $M_\rho$, naturally satisfying
the requirement stated in ref.~\tHSZ\ for the scale dependence of
a quark model ansatz for tensor TFFs.
We can therefore consider the shape of the tensor TFFs independently
of the choice of $k_T$, which we shall discuss further when
considering SDCs for the HLbL amplitude.

In fig.~\ref{fig:belle}
we show that for the singly virtual TFF, for which data have
been published by the Belle collaboration in \cite{Belle:2015oin},
there is quite good agreement for the holographic hard-wall\footnote{In \CGN,
both the hard-wall and the soft-wall model have been considered, agreeing
with our results on the singly-virtual TFF. The soft-wall model turns out
to compare less favorably with the experimental data, lying midway between
the hard-wall results and the quark model results with $\Lambda_T=M_\rho$.}
result
for $\F^T_1$ ($\F^T_3$ does not contribute in the singly virtual case).
For the normalized helicity-2 TFF, in particular
for the lowest virtualities that have been measured, there is significantly
better agreement
than that obtained with the quark model \cite{Schuler:1997yw} $\F^T_1$
with scale parameter $\Lambda_T$ set to either $M_\rho$ \cite{Hoferichter:2024bae} or $M_T$ \cite{Hoferichter:2020lap,Schuler:1997yw}.
This also holds true for the other helicities, although there the agreement
at the lowest virtualities is not as good. Like the quark model,
the holographic result fails to account for a nonvanishing helicity-0 contribution at zero
virtuality as this would require a non-zero $\F^T_2$.

In ref.\ \cite{Danilkin:2016hnh}, the Belle data have been fitted with
a quark-model scale parameter of $\Lambda_T=1.222(66)$ GeV, slightly less than $m_T$. 
This implies
a slope parameter of $2/1.222^2=1.34 \,\mathrm{GeV}^2$. While the hQCD result
provides a fit of roughly comparable quality, it involves a significantly
larger slope parameter, $1.203/M_\rho^2=2.00 \,\mathrm{GeV}^2$,
half-way to the slope parameter $2/M_\rho^2=3.33 \,\mathrm{GeV}^2$ obtained
when $\Lambda_T$ is set to $M_\rho$ as advocated in \cite{Hoferichter:2024bae}.

Unfortunately, no experimental data for the doubly virtual case are available,
which would test both $\F^T_1$ and $\F^T_3$. The doubly virtual TFFs
enter in the one-loop process $T\to e^+ e^-$, however only experimental upper limits
for the branching ratios are  currently available  for $f_2(1270)$ and $a_2(1320)$ \cite{Achasov:2000zx}.

\subsection{Asymptotic behavior}

The asymptotic behavior of tensor TFF in the cases  where one or both photon momenta are Euclidean and large, with respect to $\Lambda_{QCD}$, has been studied in ref.~\cite{Hoferichter:2020lap}, using the methods of light-cone expansion. 
Their results generalize those obtained for  pseudoscalar meson TFF, in the symmetric limit \cite{Nesterenko:1982dn,Novikov:1983jt}, and in the singly-virtual case (Brodsky-Lepage limit \cite{Lepage:1979zb,Lepage:1980fj,Brodsky:1981rp}),
showing  a milder asymptotic decrease compared to what would be obtained by VMD
with a finite number of vector meson exchanges. 

In \HS,  the following asymptotic behavior of the tensor meson TFF coefficients  $\cF^T_i$, 
written in terms of the average photon virtuality $Q$ and the asymmetry parameter $w$, has been obtained:
With\footnote{Note the different sign of the $Q^2$ defined in \cite{Hoferichter:2020lap}.}
\beq
Q^2=-\frac{q_1^2+q_2^2}{2},\;\;w=\frac{q_1^2-q_2^2}{q_1^2+q_2^2}
\eeq
the result obtained in \HS\ reads for $Q^2\to\infty$
\begin{align}
\label{BL_scaling}
 \cF_1^T(q_1^2,q_2^2)&\to\frac{4\sum_a C_a F_T^a m_T^3}{Q^4}f_1^T(w),\notag\\
 \cF_i^T(q_1^2,q_2^2)&\to-\frac{4\sum_a C_a F_T^a m_T^5}{Q^6}f_i^T(w),\qquad i\in\{2,3,4,5\},
\end{align}
with asymmetry functions
\begin{align}\label{BL_asymmetries}
 f^T_1(w)&=\frac{5(1-w^2)}{8w^6}\bigg(15-4w^2+\frac{3(5-3w^2)}{2w}\ln\frac{1-w}{1+w}\bigg),\notag\\
 f^T_2(w)&=-\frac{5}{8w^6}\bigg(15-13w^2+\frac{3(1-w^2)(5-w^2)}{2w}\ln\frac{1-w}{1+w}\bigg),\notag\\
 f^T_3(w)&=-\frac{5}{8w^6}\bigg(15-w^2-\frac{w^4+6w^2-15}{2w}\ln\frac{1-w}{1+w}\bigg),\notag\\
 f^T_4(w)&=-\frac{5}{24w^6}\bigg(45+30w-21w^2-8w^3+\frac{3(1+w)(15-5w-7w^2+w^3)}{2w}\ln\frac{1-w}{1+w}\bigg),\notag\\
 f^T_5(w)&=-\frac{5}{24w^6}\bigg(45-30w-21w^2+8w^3+\frac{3(1-w)(15+5w-7w^2-w^3)}{2w}\ln\frac{1-w}{1+w}\bigg).
\end{align}

For $\F^T_1$ and $\F^T_3$, the hQCD produces a $Q^2$ behavior in agreement with these results, however
the asymmetry functions differ:
\begin{align}
 \mathcal{F}_1&\to  -\frac{C_T m_T^3}{Q^4} \times \frac{4}{w^4}\left( 3-2w^2+\frac{3(1-w^2)}{2w}\ln\frac{1-w}{1+w}\right) \\
 \mathcal{F}_3&\to  -\frac{C_T m_T^5}{Q^6} \times  \frac{4}{w^4}\left(3 +\frac{3-w^2}{2w}\ln\frac{1-w}{1+w} \right)
\end{align}
with 
\be
C_T=\frac{\text{tr}\Q^2}{8 g_5^2 J_2(m_T z_0)}\frac{\sqrt{2/k_T}}{z_0}.
\ee

In \eqref{BL_asymmetries}, $f^T_1(w)$ vanishes in the singly virtual limit $w\to\pm 1$,
whereas the hQCD result has a finite value with
$f^T_1(1)/f^T_1(0)=5/2$. On the other hand, both $f^T_3(w)$ functions
have a minimum at $w=0$ and a logarithmic singularity for $w\to\pm 1$,
with 
\be
\lim_{|w|\to1} f^T_3(w)/f^T_3(0) = |\ln(1-|w|)|\times\begin{cases}
 \frac{105}{16} & \text{\HS} \\
 \frac{15}{4} & \text{hQCD}
\end{cases}\;.
\ee

Comparing the asymptotic ratio of $\F_3^T/\F_1^T$ in the symmetric limit we
find the same sign but a different number,
\be
\lim_{Q\to\infty} \frac{\F_3^T(-Q^2,-Q^2)}{\F_1^T(-Q^2,-Q^2)} = \frac{m_T^2}{Q^2}\times\begin{cases}
  \frac{16}{9}& \text{\HS} \\ 
  \frac23 & \text{hQCD}
\end{cases}\;.
\ee

Note that  at large momenta, $\F^T_1 T_1^{\mu\nu\alpha\beta}$ does not
dominate over the other structures, since the $T_{2\ldots 5}^{\mu\nu\alpha\beta}$ involve
two more powers of photon momenta.

\section{Tensor meson contribution to $a_\mu^{HLbL}$}
\label{sec:amuT}

The tensor meson contribution to $a_\mu^{HLbL}$ has been evaluated in ref.~\CGN\,   restricted, however, 
to the helicity-2 amplitude only, where $\F^T_1$ and $\F^T_3$ appear in the combination \HS
\be
H_{+-;+2}=H_{-+;-2}\propto \left(\F^T_1+\frac{q_1^2 q_2^2}{q_1\cdot q_2 \,m_T^2}\F^T_3\right),
\ee
if the other structure functions are set to zero (here the different sign mentioned in
footnote \ref{fn:sign} does matter).
 
In the following, we shall provide a complete evaluation, first by employing the
hQCD results for $\F^T_{1,3}$ in the formulae obtained in \cite{Hoferichter:2024fsj},
which define the pole contribution of tensor mesons in the dispersive approach, and, second, by keeping all
non-pole contribution as given by the hQCD result.
In contrast to pseudoscalar and axial vector exchange
contributions, where the two approaches lead to the same
formulae, in the case of tensor mesons they are different (cf.\ Appendix \ref{appB}). 



Furthermore, we consider two options: On the one hand,  we take the hQCD
model at face value, with one flavor-singlet tensor mode that
is implemented just like a tensor glueball in top-down
holographic models (for which radiative decays have been
worked out recently in \cite{Hechenberger:2023ljn}), but which
in the hard-wall model turns out to reproduce surprisingly well 
both mass and two-photon width of the dominant $f_2(1270)$ resonance.
On the other hand, taking the hQCD results for the TFF,
which reproduce quite well the singly virtual data for $f_2(1270)$ \cite{Belle:2015oin},
as just a model
for meson TFFs, we refit mass and two-photon width of the
lowest tensor mode to exactly match the experimental
values of the ground-state multiplet $f_2(1270)$, $a_2(1230)$, and $f_2'(1525)$.

Since the recent study of subleading HLbL contributions to $a_\mu$,
which included tensor mesons, involves a matching to pQCD results
with a separation scale of $Q_0=1.5$ GeV, we also present
a break-up of the various contributions in an IR region
defined by $Q_i\le Q_0=1.5$ GeV, a UV region $Q_i > Q_0=1.5$ GeV for all $i$, and a mixed region for the remainder.

\subsection{Pole contributions}

Our results for the pole contribution as defined by the dispersive
approach \cite{Hoferichter:2024fsj} are provided in Table \ref{tab:HWresultsdisp}.

\begin{table*}[h]
\footnotesize
    \centering
    \begin{tabular}{c|c|c|c|c|c}
    \toprule
    $M_T$ [GeV] & $\Gamma_{\gamma \gamma}$ [keV] &IR &Mixed &UV & sum $[10^{-11}]$\\
    \colrule
       1.235 ($f_2$ only) & 2.46  & $7.00-4.87=2.13$ & $0.40-0.23=0.16$   &  $0.005$ & 2.31 \\
    \hline
    \9{1.2754(8)} & \9{2.65(45)} & $6.59-4.31=2.28$ & $0.37-0.22=0.16$ & $0.004$ & 2.44(41) \\
    \9{1.3182(6)} & \9{1.01(9)} & $2.19-1.34=0.85$ & $0.12-0.07 = 0.05$ & $0.001$ & 0.90(8) \\
    \9{1.5173(24)} & \9{0.08(2)}   & $0.10-0.04=0.06$ & $0.005-0.002=0.003$ & $<0.0001$ & 0.06(2)\\
    $f_2+a_2+f_2'$ & & $8.87-5.69=3.19$ & $0.50-0.29=0.21$ & $0.005$ & 3.40(42) \\
    \hline
    1.235 ($F_\rho$-fit)  & 2.3+0.8+0.2 & $9.61-6.68=2.93$ & $0.55-0.32=0.23$ & 0.007 & 3.17 \\
    1.235 (OPE fit) & 2.6+0.9+0.2 & $10.76-7.48=3.28$ & $0.61-0.36=0.25$ & 0.008 & 3.55 \\
    \botrule
    \end{tabular}
    \caption{$a_\mu$ results for the tensor meson contributions, in units of $10^{-11}$, obtained by inserting the hQCD results for $\F^T_{1,3}$ in the formulae obtained in \cite{Hoferichter:2024fsj} within the dispersive approach, where 
    the first line corresponds to the lowest HW tensor mode with $k_T$ normalized as in \cite{Katz:2005ir}
    and identified with $f_2$ only,
    and the following lines to a refit to experimental data (written in slanted fonts to distinguish them from holographic results).
    The IR region is defined by $Q_i\le Q_0=1.5$ GeV, the UV region by $Q_i > Q_0=1.5$ GeV for all $i$. The first term of each sum is from $\overline{\Pi}_{1,2}$, while the second one is from the other 10 structure functions.
    The last two lines correspond to $k_T$ determined by matching the LSDC
    as discussed in sec.~\ref{sec:LSDC}, with fitted $F_\rho$ and OPE, respectively,
    where the lowest HW tensor mode is interpreted as a complete (degenerate) ground-state flavor multiplet.
    }
    \label{tab:HWresultsdisp}
\end{table*}

Similarly to \cite{Hoferichter:2024bae}, we observe large
cancellations between contributions from the longitudinal
pieces $\overline{\Pi}_{1,2}$ and the remaining $\overline{\Pi}_{i}$,
however in contrast to \cite{Hoferichter:2024bae}, the net result
is positive.

In Table \ref{tab:QMresults}, we show the pole contributions
obtained by using the quark model TFF \eqref{eq:FT1QM} with
$\Lambda_T$ set to either $M_\rho$ as in \cite{Hoferichter:2024bae} or $M_T$ as originally in \cite{Hoferichter:2020lap,Schuler:1997yw},
and we compare to the
hQCD result with only $\F^T_{1}$ (the latter also for $a_2$ and $f_2'$). In this case the contributions from $\overline{\Pi}_{1,2}$
are reduced, the remaining ones increased, and the net result
has a negative sign as in the quark model case. Compared
to the holographic result where $\F^T_3$ is not left out,
the absolute value is significantly larger.

\begin{table*}[h]
\footnotesize
    \centering
    \begin{tabular}{l|c|c|c|c|c}
    \toprule
     & $M_T $ [GeV]  
     &IR &Mixed &UV & sum $[10^{-11}]$\\
    \colrule
    QM($M_\rho$) & \9{1.2754(8)} & 
    $1.94-3.84=-1.89$ & $0.04-0.06=-0.02$ & $-0.001$ & $-1.91(32)$
    \\
    QM($M_T$) & \9{1.2754(8)} & 
    $5.71-11.01=-5.30$ & $0.65-0.93=-0.28$ & $-0.002$ & $-5.58(95)$
    \\
    \hline
    HW-$\mathcal{F}^T_1$ & \9{1.2754(8)} & 
    $3.44-6.81=-3.37$ & $0.18-0.25=-0.07$ & $-0.005$ & $-3.45(59)$ \\
    HW-$\mathcal{F}^T_1$ & \9{1.3182(6)} & 
    $1.11-2.21=-1.10$\ & $0.06-0.08=-0.02$ & $-0.002$ & $-1.13(10)$ \\
    HW-$\mathcal{F}^T_1$ &\9{1.5173(24)} & 
    $0.04-0.09=-0.04$ & $0.002-0.003=-0.001$ & $-0.0001$ & $-0.045(10)$ \\
    HW-$\mathcal{F}^T_1$ & $f_2+a_2+f_2'$ & 
    $4.60-9.11=-4.51$ & $0.24-0.34=-0.10  $ & $-0.007$ & $-4.62(60)$ \\
    \botrule
    \end{tabular}
    \caption{$a_\mu$ results for the tensor meson contributions, in units of $10^{-11}$, obtained by inserting the quark model TFF for $\F^T_{1}$ ($\F^T_{2,3,4,5}=0$)  in the formulae obtained in \cite{Hoferichter:2024fsj}, with
    $\Lambda_T$ set to either $M_\rho$ \cite{Hoferichter:2024bae} or $M_T$ \cite{Hoferichter:2020lap,Schuler:1997yw},
    compared to the HW result with only $\F^T_{1}$ (the latter also for $a_2$
    and $f_2'$).
    The errors in the final sum correspond to the experimental errors of the two-photon decay width used as input (as listed in tab.\ \ref{tab:HWresultsdisp}).
    }
    \label{tab:QMresults}
\end{table*}

\subsection{Full holographic contributions}

In Table \ref{tab:PoleVsNonpoleResults}, we compare
result of the reduction to the pole contribution
as defined in the optimized basis of \cite{Hoferichter:2024fsj}
to the result obtained when the complete HLbL amplitude
arising in hQCD from the exchange of tensor mesons
is employed. The latter also involves
the trace part of the metric fluctuations
as well as their longitudinal components, corresponding to
the fact that off-shell the Fierz-Pauli propagator is neither
traceless nor transverse. This gives rise to extra
tensor structures in the vertices, which in fact involve
the same two basis tensors that appear in the case of
scalar mesons
(see Appendix \ref{appB} for the details).

The first line of Table \ref{tab:PoleVsNonpoleResults}
displays the difference for the ground-state tensor meson
with model-given mass of 1.235 GeV. The
contribution to $a_\mu$ is positive in both cases,
but turns out to be more than twice
as large in the full evaluation.

\begin{table*}[h]
    \centering\footnotesize
    \begin{tabular}{c|c|r|r}
    \toprule 
    $M_T$ [GeV] & $\Gamma_{\gamma \gamma}$ [keV] & $a_\mu^\mathrm{pole}$ $[10^{-11}]$ & $a_\mu^\mathrm{full}$ $[10^{-11}]$\\
    \colrule
           1.235  & 2.46& $7.43-5.12=2.31$ & $9.33-3.24=6.09$ \\
           1.235$|_\mathrm{IR}$  & & $7.00-4.87=2.13$ & $8.64-3.24=5.40$ \\
       2.262 & 0.60& $2.17+0.57=2.74$ & $1.30-0.22=1.08$ \\
       3.280 & 1.91& $0.96+0.30=1.26$ & $0.54-0.11=0.43$ \\
       4.295 & 0.75& $0.42+0.14=0.56$ & $0.22-0.04=0.19$ \\
       5.310 & 1.75& $0.25+0.09=0.34$ & $0.13-0.02=0.11$ \\
       \hline
       [1.2--5.3] & & $11.23-4.02=7.21$ & $11.52-3.63=7.89$ \\ 
       \phantom{}[1.2--5.3]$|_\mathrm{IR}$ & & $9.62-3.67=5.95$ & $9.72-3.52=6.19$ \\
       \hline
       [1.2--10.4] & & $11.62-3.87=7.76$ & $11.71-3.66=8.05$ \\ 
       \phantom{}[1.2--10.4]$|_\mathrm{IR}$ & & $9.72-3.60=6.12$ & $9.73-3.53=6.20$ \\
       \hline
       [1.2--$\infty$] & &  &  $11.79-3.68=8.11$ \\ 
       \phantom{}[1.2--$\infty$]$|_\mathrm{IR}$ & &  & $9.74-3.53=6.21$ \\
       \hline
\botrule
    \end{tabular}
    \caption{Comparison of a full evaluation of the holographic results to the pole contribution
    as defined by the dispersive approach of ref.~\cite{Hoferichter:2024fsj}, for the first five
    tensor modes. The first term of each sum is from $\overline{\Pi}_{1,2}$, while the second one is from the other 10 structure functions. The lower part of the table shows the amounts of the first five and of the first ten tensor modes, and that of the entire tower, each also with the part from the IR region defined by $Q_i\le Q_0=1.5$ GeV.
    The sum over the pole contributions appears to converge much more slowly than the fully evaluated one. (All numbers refer to $k_T$ prior to the rematching of sec.\ \ref{sec:kTrematch} so that only the $f_2$ tower is represented.)
    }
    \label{tab:PoleVsNonpoleResults}
\end{table*}

Also shown are the $a_\mu$ contributions obtained
by evaluating excited tensor modes.
Even though the first excited tensor mode has a
mass above 2 GeV and a rather small two-photon width,
its $a_\mu$ contribution is even larger than the
ground-state tensor when only the pole contribution is kept.
In contrast, the full evaluation strongly reduces this contribution. Higher tensor modes give
smaller contributions, but they fall comparatively
slowly with mode number unless fully evaluated.
The entries of Table \ref{tab:PoleVsNonpoleResults} marked by $|_\mathrm{IR}$ 
show that the unnaturally large contributions from excited tensor mesons
obtained in the pole-only formulae
are appearing chiefly in the IR region. 

Summing over the first five modes gives rather similar results,
the difference being essentially in the
attribution to the individual modes. In the full
evaluation with non-pole terms included, the IR contribution of the 
lowest mode is already providing about 90\% of
the summed IR contribution, whereas in
the case of the pole contribution this portion is only about a third.

\subsection{Summing the infinite tower of tensor mesons}

The infinite sum over excited tensor mesons can be performed analytically using the formula
\begin{align}
    G(q^2,z,z')=\sum_n \frac{h_n(z) h_n(z')}{q^2-m_n^2},
\end{align}
where $G$ is the Green's function of the differential operator $\left({z^3}\partial_z z^{-3}{\partial_z}+{q^2}\right)$ 
with boundary conditions 
$G=0$ at $z=0$, $\partial_z G=0$ at $z=z_0$, and  $G(q^2,z,z')=G(q^2,z',z)$.
For space-like $q^2=-Q^2$, it reads
\begin{align}
\label{eq:scalbutbu}
G(q^2,z,z')
 =& -\theta(z'-z)
z^2 z'^2 I_2(Q z) \left(\frac{K_1(Q z_0) I_2(Q z')}{I_1(Q z_0)}+K_2(Q z')\right)
   \nonumber\\
   &-\theta(z-z')z^2 z'^2 I_2(Q z')
   \left(
   \frac{K_1(Q z_0) I_2(Q z) }{I_1(Q z_0)}
   +K_2(Q z)
   \right),
\end{align}
which reduces to
\be
   G(0,z,z')
 = -\frac14 \text{min}(z^4,z'^4)
\ee
at vanishing $q^2$.

It can be easily seen that each term in any $\hat{\Pi}_i$ is of the form 
$ \frac{h_n(z) h_n(z')}{q^2-m_n^2}(\frac{1}{m_n^2})^\kappa$ times $n$-independent factors (sometimes with $q^2=0$). The factors $h_n(z),h_n(z')$ come from the two $T \gamma \gamma$ vertices, while $ \frac{1}{q^2-m_n^2}(\frac{1}{m_n^2})^\kappa$  with $\kappa =0,1,2$ arises from the Fierz-Pauli propagator \eqref{eq:FP-prop}. 
For $\kappa=1,2$ we can use
\begin{align}
    G_1(q^2,z,z')=\sum \frac{1}{m_n^2} \frac{h_n(z) h_n(z')}{q^2-m_n^2}=\frac{1}{q^2}\left( G(q^2,z,z')-G(0,z,z') \right),\\
     G_2(q^2,z,z')=\sum \frac{1}{m_n^4} \frac{h_n(z) h_n(z')}{q^2-m_n^2}=\frac{1}{q^2}\left( G_1(q^2,z,z')-G_1(0,z,z') \right),
\end{align}
leading to the formulae of appendix \ref{appC} for the tensor
contributions to the $\hat{\Pi}_i$ structure
functions of the HLbL amplitude.

Another way of obtaining $\hat{\Pi}_i$ for the full tower of tensors is to start with the 5-dimensional Witten diagram for $\Pi^{\mu \nu \lambda \rho}$ and then redo the BTT projection to get $\hat{\Pi}_i$.
These two different approaches are, of course, completely equivalent and yield the same result.

\subsection{Contribution to longitudinal short-distance constraints}
\label{sec:LSDC}

\begin{figure}[t]
\centering
\includegraphics[width=0.873\textwidth]{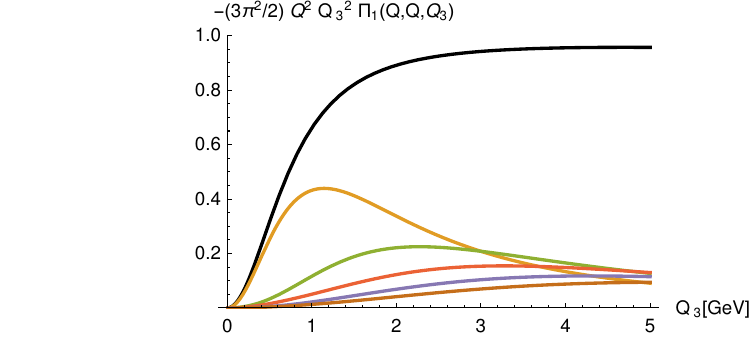}\qquad\qquad\qquad\\
\qquad\qquad\qquad\includegraphics[width=0.65\textwidth]{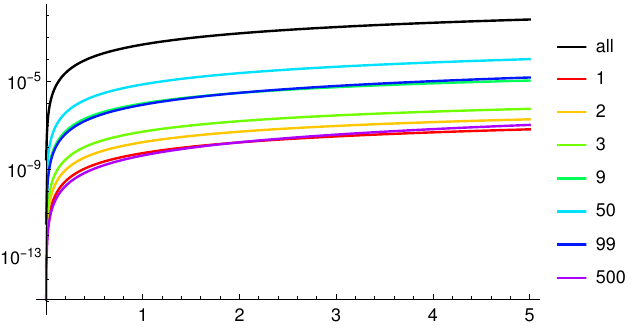}
\caption{Negligible contribution of tensor mesons to the asymmetric
longitudinal SDC: The upper panel reproduces Fig.~4 of \cite{Leutgeb:2019gbz},
which shows the contribution of the infinite tower of axial vector mesons (black
curve) saturating the MV-SDC in the
asymmetric limit $Q_3\ll Q_1\sim Q_2 \to\infty$ by plotting
$\hat\Pi_1(Q,Q,Q_3)$ for increasing $Q_3$ at $Q=50$ GeV, together with
the contribution of the first five axial vector modes (orange, green, red, purple, brown) which all tend to zero individually. The lower panel shows the
tensor contributions on a logarithmic scale, 
which could hardly be distinguished
from zero in the linear scale of the upper panel. Here the maximal contribution
is from modes with masses comparable to the chosen $Q$; all others are strongly suppressed.
\label{fig:Pi1MVSDC}}
\end{figure}

\begin{figure}[t]
\centering
\includegraphics[width=0.65\textwidth]{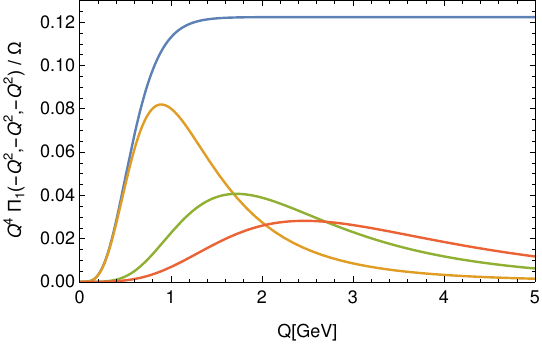}
\caption{Contribution of tensor mesons to the symmetric longitudinal
SDC: $Q^4 \,\hat\Pi_1(-Q^2,-Q^2,-Q^2)$ with full tensor bulk-to-bulk propagator (blue)
and the contributions from the first three modes (orange, green, red),
normalized to the full symmetric LSDC value $\Omega=-4/(9\pi^2)$
with $k_T$ fixed by \eqref{kTpQCD}.
\label{fig:Pi1SymSDC}}
\end{figure}

As shown by Melnikov and Vainshtein \cite{Melnikov:2003xd},
the longitudinal structure function $\bar\Pi_1\equiv 
\hat\Pi_1$ of the HLbL amplitude is subject to a SDC
following from the one-loop exact axial anomaly in combination
with the OPE of pQCD when two virtualities are much larger than
the third, $Q=Q_1 \sim Q_2 \gg Q_3$,
and the fourth set to zero,
\begin{align}
    \C_\mathrm{MV}=\lim_{Q_3\to\infty}\lim_{Q\to\infty}
    Q^2 Q_3^2 \bar\Pi_1(Q,Q,Q_3)= -\frac2{3\pi^2}.
\end{align}

Because the TFFs of meson resonances have a power-law decay in $Q^2$
and $Q_3^2$ and their propagator contributes also an inverse power of $Q_3^2$,
one needs an infinite number of resonances to have a chance to contribute to the Melnikov-Vainshtein SDC (MV-SDC).

In \cite{Leutgeb:2019gbz,Cappiello:2019hwh,Leutgeb:2021mpu} it was
shown that the infinite tower of axial vector mesons, which is unavoidable in hQCD models and whose photon interactions are determined
by the Chern-Simons action implementing the axial anomaly, indeed
saturates the MV-SDC whenever the 5-dimensional geometry is
asymptotically AdS, yielding
\begin{align}
    \C_\mathrm{MV}^\mathrm{A}=-\left(\frac{g_5}{2\pi}\right)^2\frac1{\pi^2}\int_0^\infty d\xi\, \xi [\xi K_1(\xi)]^2=-\frac2{3\pi^2}\left(\frac{g_5}{2\pi}\right)^2,
\end{align}
where $g_5=2\pi$ when the vector current two-point function is
matched to the leading-order pQCD result \eqref{PiVas}.
As shown in \cite{Leutgeb:2021mpu}, in hQCD models which also have an 
infinite tower of pseudoscalar mesons,\footnote{The chiral Hirn-Sanz model \cite{Hirn:2005nr} has only the ground-state Goldstone bosons.} it is still only the axial vector mesons which realize the MV-SDC.

On the other hand,
the symmetric short-distance limit of $\bar\Pi_1$ reads
\cite{Melnikov:2003xd,Colangelo:2019lpu,Colangelo:2019uex,Bijnens:2020xnl,Bijnens:2021jqo}
\begin{align}
    \C_\mathrm{sym}=\lim_{Q\to\infty}
    Q^4 \bar\Pi_1(Q,Q,Q)=-\frac4{9\pi^2},
\end{align}
but the axial sector of hQCD models yields a smaller result,
\begin{align}\label{Pi1AVsymlimit}
\C_\mathrm{sym}^\mathrm{A}&=-\left(\frac{g_5}{2\pi}\right)^2\frac1{\pi^2}
\int_0^\infty d\xi  
\,\xi [\xi K_1(\xi)]^3 \nonumber\\
&=-0.8122\times\frac{4}{9\pi^2}\left(\frac{g_5}{2\pi}\right)^2.
\end{align}

Because also tensor mesons form an infinite tower of resonances in
hQCD models, they can potentially contribute to the MV-SDC as well
as to the symmetric SDC for $\bar\Pi_1$.
With two of the momenta set equal, the complete expression
for $\bar\Pi_1$ given in \eqref{Pi1TGreen} can be simplified
after partial integrations to
\begin{align}\label{Pi1QQQ3}
      \bar{\Pi}_1&(Q,Q,Q_3)= -\frac{4}{k_T} \left(\frac{\text{tr}\Q^2}{g_5^2}\right)^2 \int_0^{z_0} \frac{dz dz'}{z z'}\J(Q,z) \J(Q,z') \frac{ \partial_{z'}\J(Q_3,z')}{Q_3^2} \partial_{z'}G(0,z,z')
\end{align}
with $G(0,z,z')=-\frac{1}{4}\text{min}(z^4,z'^4)$.

In the MV-limit $Q_3\ll Q\to\infty$, this result decays faster than
$1/Q^2$ and so does not contribute to the MV-SDC. This is illustrated
in fig.~\ref{fig:Pi1MVSDC} for $Q=50$ GeV and increasing $Q_3$,
where the tower of axial vector mesons builds up the correct
asymptotic behavior, while the tensor meson contributions remain
negligible.

However, in the symmetric limit $Q_3=Q\to\infty$, there is a
nonvanishing contribution reading\footnote{Here both $\F^T_1$
and $\F^T_3$ contribute at the same order. With only $\F^T_1$,
the number $-0.15285$
in \eqref{Pi1QQQ3num} would be reduced to $-0.097573$, but
the simplification from partial integrations leading to \eqref{Pi1QQQ3}
would not occur.}
\begin{align}\label{Pi1QQQ3num}
Q^4\bar{\Pi}_1(Q,Q,Q)\to&
    \frac{4}{k_T} \left(\frac{\text{tr}\Q^2}{g_5^2}\right)^2\int_0^\infty d\xi K_1(\xi) 
    \int_0^\xi d\xi' \xi'^3  K_1(\xi')\, \partial_{\xi'}\left[\xi' K_1(\xi')\right],
    \nonumber\\
&=\frac{4}{k_T} \left(\frac{\text{tr}\Q^2}{g_5^2}\right)^2 \times (-0.15285). 
\end{align}
This is $12.23 \%$ of the OPE result for $N_c=N_f=3$
when following \cite{Katz:2005ir} $k_T$ is matched through the energy-momentum tensor two-point function by \eqref{kTpQCD}.
Axial vector mesons already give 81.22\%, so the symmetric
SDC is now realized at the level of 93.45\%.
Reducing $k_T$ by leaving out the quark contribution
in the energy-momentum tensor two-point function in \eqref{kTpQCD}
and thus keeping only the leading part in a large-$N_c$ limit
would increase this further to 98.2\%. However,
this is obviously fortuitous, because the large-$N_c$ counting
is still not correct for tensor mesons forming flavor multiplets
as we shall now argue.

\subsection{Rematching $k_T$ with correct large $N_c$ behavior}
\label{sec:kTrematch}

The short-distance limits of the HLbL amplitude are all
proportional to $N_c \text{tr}\Q^4$, 
which is also the case for the axial-vector meson contributions,
but  the above result for the tensor mesons
approaches $N_c^0 (\text{tr}\Q^2)^2$ because
$g_5^2=12\pi^2/N_c$.

For the correct large-$N_c$ scaling, $k_T$ should be proportional
to a single power of $N_c$, as it would be when only the quark
part of the energy-momentum tensor two-point function in \eqref{kTpQCD}
was retained. Moreover, the individual tensor mesons
of a flavor multiplet should contribute according to the
sum
$\sum_{a=0,3,8}(\text{tr}\Q^2 T^a)^2=\frac12 \text{tr}\Q^4$.

The above result for a single tensor mode is instead
characteristic of a tensor glueball.
However, in AdS/QCD, SU($N_f$)-symmetric flavor multiplets of tensor mesons
can still be modeled by the universal tensor mode, if
their asymptotic two-point functions have the form given in
\eqref{kTpQCD} apart from overall factors.
The correct $N_c$ and $N_f$ dependence follows if $k_T$
is forced to match the deficit of the symmetric SDC
left by the axial-vector mesons.

With $g_5=2\pi$ so that the OPE of the vector correlator is matched,
the value of $k_T=5/(16\pi^2)$ following from \eqref{kTpQCD}
for $N_c=N_f=3$ needs just to be reduced by a factor 1.536,
leading to a simple overall enhancement of all results
given above by the same factor.

In \cite{Leutgeb:2022lqw} it was proposed to take into
account typical next-to-leading order gluonic corrections at
moderately high energies by using
a reduced $g_5$, for definiteness chosen as
$g_5^2=0.894(2\pi)^2$ by fitting $F_\rho$ instead of the
vector correlator OPE.
This enhances
the contribution from tensor mesons by a factor $1/0.894^2$ relative
to the axial-vector contributions 
so that an extra factor of 1.097 suffices
to keep the correct ratio between symmetric and MV SDC.
In this latter scenario an overall rescaling of the above tensor
contributions by a factor of 1.373 in place of 1.536 is needed.

For our final estimates we shall adopt the latter for a central
value, the larger result obtained by fitting the vector-correlator
OPE as upper value, and the result with the original choice of $k_T$
as lower one.

Ref.~\cite{Katz:2005ir} has also considered the ratio $\Gamma(f_2\to\pi\pi)/\Gamma(f_2\to\gamma\gamma)$ with the result that the model for the axial sector of ref.~\cite{Erlich:2005qh}
yields only about 25\% of the experimental value. While this ratio is independent of the normalization of the tensor mode, it changes when the tensor mode is distributed over a flavor multiplet as opposed to a single glueball-like meson. Assuming for example that $f_2(1270)$ is ideally mixed, involving only the first two flavors, this ratio becomes enhanced
by a factor $(6/5)^2=1.44$ due to the different factor $\text{tr}(\Q^2 T^a)$ in the two-photon amplitude, mitigating somewhat the numerical deficit noted in ref.\ \cite{Katz:2005ir}.

\subsection{Final results}


The last two lines of table \ref{tab:HWresultsdisp} contain the results
of the rematching of $k_T$ as described in sec.~\ref{sec:kTrematch},
corresponding simply to a multiplicative factor of 1.373 and 1.536
for the reduced ($F_\rho$-fitted) and the full OPE-fit of the LSDCs,
with respect to the original choice \eqref{kTpQCD} of ref.~\cite{Katz:2005ir}.
While the latter reproduces very well mass and two-photon rate
of $f_2(1270)$, the former should represent the complete ground-state
multiplet $f_2(1270)$, $a_2(1320)$, and $f_2'(1525)$ in a
flavor-symmetric approximation. Assuming ideal mixing, the full
two-photon width would be distributed as
$2.3+0.8+0.2$ and $2.6+0.9+0.2$ keV, for $F_\rho$-fit and OPE fit,
respectively, which roughly reproduces
the experimental pattern of $2.65(45),1.01(9),0.08(2)$ keV
for $f_2,a_2,f_2'$, and together brackets the overall two-photon
coupling.
Fitting the experimental values individually with adjusted $k_T$
values, the holographic results for the pole contribution
add up to a value of $a_\mu^{T_1\, \mathrm{pole}}\times 10^{11}=3.4(4)$, which is in between
the results 3.17 and 3.55 obtained with the effective $k_T$ values as determined
by the $F_\rho$ and OPE fits.

With the $F_\rho$-fitted result as central value, we summarize
the holographic result for the pole contribution of the ground-state
tensor multiplet as
\be
a_\mu^{T_1\, \mathrm{pole}}\times 10^{11}=3.17_{-0.86}^{+0.38}\quad [\text{IR:}\; 2.93_{-0.80}^{+0.35}]
\ee
where the latter value corresponds to the contribution from $Q_i\le Q_0=1.5$ GeV.
The effect of including non-pole contributions can be obtained by simple rescalings from table \ref{tab:PoleVsNonpoleResults}, which leads to the significantly larger
values
\be
a_\mu^{T_1 \,\mathrm{full}}\times 10^{11}=8.36_{-2.27}^{+0.99}\quad [\text{IR:}\; 7.41_{-2.01}^{+0.88}].
\ee

Experimental data on two-photon couplings exist also for the first
excited tensor modes $f_2(1565)$ and $a_2(1700)$, reading
0.70(14) and 0.30(6) keV, respectively.
Their masses are overestimated by the hard-wall holographic
result of 2262 MeV, but the rescaled two-photon width of the latter of
0.82 and 0.92 keV, for $F_\rho$ and OPE fits, resp., are comparable
to their combined effects. 
Our hQCD predictions for their $a_\mu$-contributions are
\begin{align}
    a_\mu^{T_2 \,\mathrm{pole}}\times 10^{11}=3.76_{-1.02}^{+0.44}\quad 
    &[\text{IR:}\; 3.30_{-0.90}^{+0.39}],\\
    a_\mu^{T_2 \,\mathrm{full}}\times 10^{11}=1.48_{-0.40}^{+0.18}\quad &[\text{IR:}\; 0.81_{-0.22}^{+0.10}].
\end{align}
Note that this is now much larger when only the pole term is kept, which
even surpasses the contribution from the $n=1$ mode.

As table \ref{tab:PoleVsNonpoleResults} shows, summing over the first few
modes gives rather similar results for pole contributions and full evaluations.
Summing over the entire tower finally gives
\begin{align}\label{amufulltower}
    a_\mu^{T}\times 10^{11}=&11.1_{-3.0}^{+1.3}\nonumber\\
    &[\text{IR:}\; 8.5_{-2.3}^{+1.0}]\nonumber\\
    &[\text{Mixed:}\; 1.9_{-0.5}^{+0.2}],
\end{align}
where we have also given the contributions from the Mixed region.
The UV regions contributes around $0.7\times 10^{-11}$, which
is comparatively negligible, but combined with the 
contributions from axial vector mesons in the ``best-guess'' hard-wall model of ref. \cite{Leutgeb:2022lqw},
this adds up
to a value of $6.3(7)\times 10^{-11}$, bringing it into perfect agreement
with the dispersive result for subleading contributions \cite{Hoferichter:2024bae,Hoferichter:2024vbu}, which reads $6.2_{-0.3}^{+0.2}\times 10^{-11}$.

\section{Conclusions}

The surprisingly large positive contributions to $a_\mu$
that we have obtained in hQCD for tensor mesons are in stark
contrast to the negative result obtained in \cite{Hoferichter:2024bae,Hoferichter:2024vbu}
using a simple quark model ansatz,
$-2.5\times10^{-11}$ from the IR region.
The contributions there of further intermediate states
have been determined as $+2.0\times10^{-11}$ and
are well matched by the contributions of
axial vector mesons and excited pseudoscalars
in the hQCD model of \cite{Leutgeb:2022lqw,Leutgeb:2024rfs} ($+2.3\times10^{-11}$
for the $F_\rho$-fit).
The additional result from the full tower of tensor mesons in the IR
region as given in \eqref{amufulltower}, if it was to replace
the quark model result adopted in \cite{Hoferichter:2024bae,Hoferichter:2024vbu}\footnote{In
\cite{Hoferichter:2024bae,Hoferichter:2024vbu} only the pole contributions from
the ground-state tensor multiplet are taken into account explicitly; the low-energy
contributions from the remaining tensor modes would be booked as ``effective poles'',
which at present is only done in the pseudoscalar and axial vector sector, however.}
would raise the final result of \cite{Hoferichter:2024bae,Hoferichter:2024vbu}
from 101.9(7.9) to about $113\times 10^{-11}$.
This would then lead to a better agreement with
recent lattice results, which all are higher: $109.6(15.9)$, $124.7(14.9)$, and $125.5(11.6)\times10^{-11}$ for
the Mainz group \cite{Chao:2021tvp,Chao:2022xzg}, the RBC-UKQCD group \cite{Blum:2023vlm}, and the BMWc lattice group \cite{Fodor:2024jyn}, respectively.

The data that are presently available for the tensor TFFs
compare remarkably well with the hQCD results as shown in
fig.\ \ref{fig:belle}, while the quark model as used
in \cite{Hoferichter:2024bae,Hoferichter:2024vbu} is severely
underestimating the experimental result even at the smallest available
virtualities. Using a larger scale than $M_\rho$ as done
originally in the fits of ref.~\cite{Danilkin:2016hnh} would
in fact give much more strongly negative contributions. As we have seen,
the positive result obtained in the hQCD case
is due to the extra structure function $\F_3^T$ which contributes
only to the doubly virtual case where data are not yet available.
It would thus be very desirable to be able to 
test the holographic prediction
for the $T\to\gamma^*\gamma^*$ amplitude, 
away from the singly virtual limit.

From a theoretical point of view, the presence of $\F_3^T$
in the holographic approach is an unavoidable consequence
of the five-dimensional nature of the flavor gauge field Lagrangian,
involving $F_{\mu z}$-components on a par with purely four-dimensional
components. Dropping the former would in fact lead to a complete decoupling
of tensor mesons from pions in a simple hQCD model such as the
Hirn-Sanz model \cite{Hirn:2005nr}. Within the 
model of \cite{Erlich:2005qh}, the $f_2\to\pi\pi$
rate has already been estimated in \cite{Katz:2005ir} and found
to be of roughly the right order of magnitude, but significantly too small.
With our different prescription 
in sec.\ \ref{sec:kTrematch}
this is in fact coming somewhat closer to experimental data.

Despite these encouraging results that are obtained with a minimal
set of free parameters, there remain also some open theoretical questions.

As we have discussed above, the light-cone expansion results of
ref.\ \cite{Hoferichter:2020lap} generalizing the Brodsky-Lepage approach  \cite{Lepage:1979zb,Brodsky:1981rp} involve all structure functions,
but differently than the hQCD results imply---in contrast to the
case of axial vector mesons, where there is full agreement.
It would also be very interesting to explore the remaining SDCs
on the HLbL amplitude which have been obtained in \cite{Colangelo:2019uex,Colangelo:2019lpu,Bijnens:2020xnl,Bijnens:2021jqo}.
This certainly calls for further studies which we intend
to pursue in follow-up work.

However, the numerical contribution of tensor mesons contributions to $a_\mu^\mathrm{HLbL}$ is
dominated by the behavior of the tensor TFFs at low energies, where we found
$\F^T_3$ to be equally important as $\F^T_1$ although only the latter of the two
contributes to the real and singly virtual cases.
A complete description will eventually also require $\F^T_{2,4,5}$, although
in the new optimized basis introduced in \cite{Hoferichter:2024fsj} they
generically lead to kinematical singularities when employed together with $\F^T_1$.
In the holographic model, this issue does not arise, while
$\F^T_1$ and $\F^T_3$ appear at a comparably important
level as being associated with kinetic and mass terms of vector mesons to whose
energy-momentum contributions the tensor mode couples.

\section*{Note Added}

After completion of the above work,
Ref.~\cite{Estrada:2025bty} has presented a calculation
of tensor meson contributions to $a_\mu^\mathrm{HLbL}$
in resonance chiral theory with a tensor-vector-vector
interaction giving rise to only $\F^T_1$. After
matching to OPE and experimental values for the
two-photon coupling with two-vector meson resonances,
results were obtained that agree completely with
the results in Table \ref{tab:QMresults} where we
show the effect of omitting $\F^T_3$ and then
matching the holographic TFF with masses and
two-photon couplings of $f_2,a_2,f_2'$.
In the appendix we discuss how the holographic results
are related to resonance chiral theory, and more generally
how the holographic result can be understood in terms
of four-dimensional actions and an expansion in vector meson modes.

\begin{acknowledgments}
We would like to thank Martin Hoferichter and Peter Stoffer for very useful discussions. 
This work was funded in part by the Austrian Science Fund (FWF), grant-DOI \url{https://www.doi.org/10.55776/PAT7221623}.  L.C. acknowledges the support of the INFN research project
ENP (Exploring New Physics).
\end{acknowledgments}

\appendix

\section{Mode decomposition of the tensor transition form factors 
and relation to resonance chiral theory}
\label{sec:modedecomposition}

In the following we discuss a mode decomposition of the tensor TFFs $\F^T_1$ and
$\F^T_3$ as obtained in
hQCD, eqs.~\eqref{F1} and \eqref{F3}, and how those (and approximations thereof using a finite number of vector meson resonances)
would be obtained in a 4-dimensional formulation.

The hQCD result involves the bulk-to-boundary propagator $\J(z,q^2)$
which for $z>0$ can be expanded in vector-meson modes\footnote{In the appendices, we switch our notation to $q^2=-Q^2$ and write $\J(z,q^2)$ in lieu of of $\J(Q,z)$.}
\begin{align}
\label{eq:Jmode}
    \J(z,q^2)=\sum\n \frac{F_n v_n(z)}{q^2-m_n^2},
\end{align}
where the $v_n(z)\equiv g_5\psi_n(x)$ are the vector-meson mode functions 
given explicitly in \eqref{eq:psin}.
They obey 
\begin{align}\label{eq:vecmodeq}
z\partial_z (z^{-1} \partial_zv_n)+m_n^2 v_n=0,
\end{align}
vanish at $z=0$ and satisfy Neumann boundary conditions in the infrared with normalization condition $g_5^{-2} \int\dz\, z^{-1}v_n(z)^2=1$. The decay constants $F_n$ are defined by 
\be
F_n=-g_5^{-2} \lim_{\varepsilon \rightarrow 0} \left(\varepsilon^{-1} \partial_z v_n(\varepsilon)\right),
\ee
and have mass dimension $2$.

The TFF $\F^T_1$ can then be simply expressed as
\begin{align}\label{eq:FT1modesum}
\F^T_1(q_1^2,q_2^2)/m_T=-\QQ \sum_{mn} c_{mn}^{(1)}F_m F_n \frac1{q_1^2-m_m^2}\frac1{q_2^2-m_n^2}
\end{align}
with
\begin{align}\label{eq:cmn1}
    {c}^{(1)}_{mn}=\frac{1}{g_5^2} \int\dz \frac{h(z)}{z}v_m(z) v_n(z).
\end{align}
However, in $\F^T_3$, which involves $\partial_z\J/q_i^2$, one
cannot use \eqref{eq:Jmode} directly, because the derivative $\partial_z$
must not be interchanged with the infinite sum:
\begin{align}
\label{eq:illegalsumrel}
    \partial_z \J(z,q^2) \not\equiv\sum_n \frac{F_n v_n'(z)}{q^2-m_n^2};
\end{align}
while the left-hand side is well-defined, the right-hand side in fact diverges.

However, using the sum rule
\begin{align}
\sum\n \frac{F_n v_n(z)}{m_n^2}=-1,
\end{align}
one can obtain an alternative version for $\J(z,q^2)$ that holds for all $z$,
\begin{align}
\label{eq:J-1modes}
    \J(z,q^2)=1+\sum\n \frac{q^2}{m_n^2} \frac{F_n v_n(z)}{q^2-m_n^2};
\end{align}
from which one gets
  \begin{align}
\label{eq:Jpmode}
    \frac{\partial_z\J(z,q^2)}{q^2}=\sum\n \frac{1}{m_n^2} \frac{F_n v_n'(z)}{q^2-m_n^2},
\end{align}
and this leads to a well-defined mode expansion of $\F^T_3$:
\begin{align}\label{eq:FT3modesum}
    -\F^T_3(q_1^2,q_2^2)/m_T^3 = \QQ \sum_{n,m} c^{(3)}_{mn}\frac{F_n }{m_n^2}\frac{F_m }{m_m^2} \frac{1}{q_1^2-m_n^2}\frac{1}{q_2^2-m_m^2}
\end{align}
with
\begin{align}\label{eq:cmn3}
{c}^{(3)}_{mn}=\frac{1}{g_5^2} \int\dz \frac{h(z)}{z}v'_m(z) v'_n(z).
\end{align}
This is regular at the origin $q_i^2=0$ irrespective of the number of modes one chooses to work with; there are also no issues of convergence of the infinite sum for the holographic result.

However, while at low $q^2$, the decomposition \eqref{eq:J-1modes} restricted to a finite number
of modes approximates the bulk-to-boundary propagator $\J(z,q^2)$ better than
\eqref{eq:Jmode}, the opposite holds at large $q^2$. 
With a finite number of modes, the correct short-distance power-law behavior is
easily achieved for $\F^T_1\sim 1/q^4$, but $\F^T_3\sim 1/q^6$ as ensured
by the holographic result would in general require corrections by hand.

\subsection{Holographic realization of VMD}

In order to recast the hQCD model of tensor-photon interactions
as a four-dimensional
resonance model, we decompose the 5D vector gauge field into a normalizable part and a non-normalizable part
\begin{align}
    \label{eq:vectordecomp}
    V_{\mu}(z,x)= \bar{V}_{\mu}(z,x)+ \sum_n v_n(z) V^n_\mu(x),\quad
    V_\mu^n(x)\equiv v_\mu^{(n)a}t^a.
\end{align}
The non-normalizable part $\bar{V}_{\mu}(z,x)$ obeys $\bar{V}_{\mu}(0,x)=a_\mu(x)\Q$, and the modes $v_n(z)$ go to zero at the boundary and form a basis for normalizable functions. The choice of $\{v_n(z)\}$ 
as the eigenfunctions appearing in \eqref{eq:vecmodeq} leads to canonical 4D massive vector Proca action for each $V_\mu^n(x) $.
There is, however, still considerable freedom in choosing $\bar{V}_{\mu}(z,x)$. The difference between any two such choices vanishes at $z=0$ and can therefore be decomposed in eigenfunctions $\{v_n(z)\}$. By then performing a field redefinition of $V^n_\mu(x)$ it is seen that the two choices are completely equivalent
(cf.\ the discussion of VMD in hQCD in \cite{Sakai:2005yt}).
A popular choice in holography is to take the non-normalizable part to obey the linear 5D EOM $\partial_M F^{MN}=0$, in the $V_z=0$ gauge. Upon dimensionally reducing to four dimensions, the Lagrangian will contain terms that are nonlocal in $a_\mu(x),V^n_\mu(x)$.
This is less desirable from a 4D perspective, therefore we choose a different option for the non-normalizable part:
\begin{align}
\label{eq:locsplit}
    \bar{V}_{\mu}(z,x) \equiv \Q \,a_\mu(x).
\end{align}
The quadratic part of the 4D Lagrangian in the vector sector, i.e., with tensor 
modes neglected, which follows from the gauge kinetic term in 5D then reads
\begin{align}
\label{eq:locaction}
    S_{4D}&= \sum_n S_{\text{Proca}}[V^n_\mu(x)]+ \sum_n\text{tr}(\Q t^a) \frac{F_n}{m_n^2} \int d^4x F_{\mu \nu}^{a\,n}(x) f^{\mu \nu}(x)\nonumber\\&+ \frac{\text{tr}(\Q^2)}{2 g_5^2}\log(\varepsilon/z_0) \int d^4x f_{\mu \nu}(x)f^{\mu \nu}(x)
\end{align}
with $f_{\mu \nu}=\partial_\mu a_\nu-\partial_\nu a_\mu$ and the Proca Lagrangian being given by $\mathcal{L}_{\text{Proca}}= -\frac{1}{4}(F^a_{\mu \nu})^2+ \frac{m_V^2}{2}(V_\mu^a)^2$ (for the field $V^n$, the mass is $m_n^2$). This action is supplemented by a holographic counterterm $S_{ct}\sim \log(\varepsilon \mu) \int \sqrt{\gamma} F_{\mu \nu}(\varepsilon,x)F^{\mu \nu}(\varepsilon,x)$ where $\gamma$ is the induced metric on the $z=\varepsilon$ surface which effectively changes $\log(\varepsilon/z_0)\rightarrow-\log(\mu z_0)$, thereby canceling the divergence.

\subsection{Four-dimensional formulation with Proca fields}

Dimensionally reducing the 5D $TVV$ interactions that generate the $\F_1$ TFF yields
\begin{align}\label{eq:STVVFT1}
    S_{TVV}|_{\F_1^T}&= c_{\gamma \gamma} \int d^4x \; h^{\alpha \beta}(x)f_{\alpha \mu}(x)f_{\beta \nu}(x) \eta^{\mu \nu}+ \sum_n c_{\gamma n}^a \int d^4x \;  h^{\alpha \beta}(x) f_{\alpha \mu}(x)F^{a \;n}_{\beta \nu}(x) \eta^{\mu \nu}\nonumber \\&
    \quad + \frac12 \sum_{mn} c^{(1)}_{mn} \int d^4x \; h^{\alpha \beta}(x)F^{a \;m}_{\alpha \mu}(x)F^{a\; n}_{\beta \nu}(x) \eta^{\mu \nu}
\end{align}
with $c^{(1)}_{mn}$ as defined in \eqref{eq:cmn1} and
\begin{align}
    c_{\gamma \gamma}&= \frac{\text{tr}(\Q^2)}{g_5^2} \int \frac{dz}{z} h(z), \\
    c^a_{\gamma n}&=2\frac{\text{tr}(\Q t^a)}{g_5^2} \int \frac{dz}{z} h(z) v_n(z) .
\end{align}
The 5D interaction responsible for the $\F_3^T$ TFF yields
\begin{align}\label{eq:STVVFT3}
  S_{TVV}|_{\F_3^T}=- \frac12 \sum_{mn} c^{(3)}_{mn} \int d^4x\; h^{\alpha \beta}(x)V_\alpha^{a \;m}(x)V_\beta^{a \;n}(x)
\end{align}
with $c^{(3)}_{mn}$ as defined in \eqref{eq:cmn3}.

Upon calculating $\F_1^T$ via the mode decomposition, we find
\begin{align}
\label{eq:F1frommodes}
   - \F_1^T/m_T= c_{\gamma \gamma}+\sum_n c_{\gamma n}^a \text{tr}(\Q t^a) \frac{F_n}{m_n^2}\left( \frac{q_1^2}{q_1^2-m_n^2}+\frac{q_2^2}{q_2^2-m_n^2} \right)  \nonumber \\
   +\sum_{mn} 2 \text{tr}(\Q^2) \frac{q_1^2}{m_n^2}\frac{q_2^2}{m_m^2}c^{(1)}_{mn} \frac{F_n}{q_1^2-m_n^2}\frac{F_m}{q_2^2-m_m^2}
\end{align}
which upon inserting the expressions for $c_{\gamma \gamma}, c^a_{\gamma n}, c^{(1)}_{mn}$ agrees precisely with what one would get from inserting the decomposition
\eqref{eq:J-1modes}
directly into the holographic formula for the TFF \eqref{F1}.

Note that, in this formulation, the couplings responsible for the $\F^T_3$ TFF
are apparently \textit{lower} in derivatives than the ones involved in $\F^T_1$.
It is only thanks to the holographic sum rule
\begin{align}
\sum\n \frac{F_n v_n(z)}{m_n^2}=-1,
\end{align}
that one can combine the direct coupling of the tensor meson to two photons
and the couplings involving a mixing of the photon with vector mesons
in the infinite mode sum with the help of
\begin{align}
1+\sum\n \frac{q^2}{m_n^2} \frac{F_n v_n(z)}{q^2-m_n^2}=
\sum\n \frac{F_n v_n(z)}{q^2-m_n^2}=\J(z,q^2),
\end{align}
which effectively replaces $q_1^2$ and $q_2^2$
in the mode sum representation of the amplitude $\mathcal{M}|_{\F_1^T}$
by the mass of the individual tensor mesons, thereby returning
to the representation \eqref{eq:FT1modesum}
which renders manifest the \textit{complete} VMD of hQCD.
Without this rewriting, a truncation to a finite number of modes
would not automatically give a
short-distance power-law behavior of $\F^T_1\sim 1/q^4$ 
and would have to be added as a constraint by hand.

For the $\F_3$ TFF we immediately recover \eqref{eq:FT3modesum}.
As remarked above, here a truncation 
to a finite number of modes
generally has the problem
that the subleading short-distance behavior needs to be imposed as an
additional constraint.

\subsection{Antisymmetric tensor representation of vector mesons}

There is another formulation for vector mesons that uses antisymmetric tensor fields $V_{\mu \nu}^n$ instead of Proca fields, which is employed in resonance chiral theory
\cite{Gasser:1983yg,Ecker:1988te}, and in this formulation 
the latter problem
is effectively circumvented by the substitution
\begin{align}
\label{eq:procaantisymrel}
    V_\mu^n= \frac{1}{m_n} \partial^\nu V^n_{\nu \mu},\quad V_{\mu\nu}=-V_{\nu\mu}.
\end{align}
Note, however, that this relation can only hold on-shell, since only then $\partial_\mu V^\mu_n=0$.
The coupling to photons in this formulation reads
\begin{align}
    S^{\text{antisym.}}_{\gamma V}=\sum_n\text{tr}(\Q t^a)\frac{F_n}{m_n}\int V^{a \;\mu \nu}_n(x) f_{\mu \nu}(x) d^4x.
\end{align}

The propagator for $V_n^{\mu \nu}$ is given by
\begin{align}
    \langle V_n^{\mu \nu}(x) V_n^{\rho \sigma}(y) \rangle = &\frac{1}{m_n^2} \int \frac{d^4q}{(2 \pi)^4} e^{-iq(x-y)} \frac{-i}{q^2-m_n^2} \nn\\
    &\times\bigg( \eta^{\mu \rho} \eta^{\nu \sigma}(-q^2+m_n^2) +\eta^{\mu \rho}q^ \nu q^\sigma -\eta^{\mu \sigma}q^\nu q^\rho -(\mu \leftrightarrow \nu)\bigg).
\end{align}
Now the $\F^T_1$ TFF follows from ultralocal (i.e., non-derivative)
couplings 
\begin{align}
    S_{4D}|_{\F^T_1}=\sum_{m,n}m_m m_n {c}^{(1)}_{mn} \int d^4x\, h^{\alpha \beta}(x) V^n_{\alpha \kappa}(x)V_{\beta \lambda}^{m }(x)\eta^{\kappa \lambda}
\end{align}
which using
\begin{align}
\partial_\rho^y \langle V_n^{\mu \nu}(x) V_n^{\rho \sigma}(y) \rangle = - \int \frac{d^4q}{(2 \pi)^4} e^{-iq(x-y)} \frac{1}{q^2-m_n^2} \left(q^\mu \eta^{\sigma \nu}-q^\nu \eta^{\sigma \mu}\right)
\end{align}
leads to the mode sum for $\F_1^T$ as given in \eqref{eq:FT1modesum}.

There is now also a local coupling (neglected in \cite{Estrada:2025bty}) that produces a nonzero and regular $\F^T_3$,
\begin{align}
 S_{4D}|_{\F^T_3}=    \sum_{m,n} \frac{c^{(3)}_{mn}}{m_n m_m}\int d^4x \; h^{\mu \nu}(x)\partial^\rho V^n_{\rho \mu}(x)\partial^\sigma V^m_{\sigma \nu}(x).
\end{align}
This coupling is just what one would get if one were to insert \eqref{eq:procaantisymrel} into \eqref{eq:STVVFT3}. 

Since 
\begin{align}
  \partial_\mu^x \partial_\rho^y  \langle V_n^{\mu \nu}(x) V_n^{\rho \sigma}(y) \rangle= \int \frac{d^4q}{(2 \pi)^4} e^{-iq(x-y)} \frac{-i}{q^2-m_n^2}(q^2 \eta^{\nu \sigma}-q^\nu q^\sigma)
\end{align}
one obtains the mode expansion for $\F^T_3$ as given in \eqref{eq:FT3modesum},
which is regular at $q_i^2=0$, irrespective of whether the sum over vector modes
is finite or infinite, 
thus contributing only in the doubly virtual case.
However, the correct short-distance behavior
$\F^T_3\sim 1/q^6$ would need to be corrected by a constraint when the
holographic result was replaced by a finite sum.

\subsection{Relation to minimal models}

In the so-called minimal model of Ref.~\cite{Mathieu:2020zpm},
an ultralocal (non-derivative) coupling of tensor modes and two vector mesons
in Proca field representation as in
\eqref{eq:STVVFT3}
is combined by assuming an ultralocal photon-vector-meson coupling
\begin{align}\label{eq:Va}
    S_{\gamma V}^\mathrm{min}= \sum_n F_n \int d^4x \; V_n^\mu(x) a_{\mu}(x),
\end{align}
where the missing gauge invariance of this term is corrected
by enforcing transversality by adding a transverse projector to the photon lines
by hand.

This leads to a TFF $\F^T_3$ which is singular at $q_i^2=0$,
thus yielding a finite contribution to the two-photon decay rate of tensor mesons,
instead of through $\F^T_1(0,0)$.

A manifestly gauge-invariant coupling can be obtained from \eqref{eq:Va} by
adding terms proportional to the linear equations of motion for the Proca fields
according to
\begin{align}\label{eq:SpgV}
    S'_{\gamma V}&=\sum_n F_n \int d^4x \; V_n^\mu(x) a_{\mu}(x) - \sum_n \frac{F_n}{m_n^2} \int d^4x \;\left(\partial_\mu F_n^{\mu \nu}(x)+m_n^2 V_n^{\nu}(x)\right)a_{\nu}(x) \nn\\
    &=\sum_n \frac{F_n}{m_n^2}\int d^4x\, \partial_\mu F_n^{\mu \nu}(x)a_{\nu}(x).
\end{align}
The effect of the additional terms is to introduce a contact term to the
Proca propagator which renders it transverse also off-shell
\begin{align}
    G_{P_n}^{\mu\nu}(q) &= \frac{-i}{q^2-m_n^2}\left(\eta^{\mu \nu}-\frac{q^\mu q^\nu}{m_n^2}\right) \nonumber\\
    &\to
    \frac{-i}{q^2-m_n^2}\left(\frac{q^2}{m_n^2}\eta^{\mu \nu}-\frac{q^\mu q^\nu}{m_n^2}\right)= G^{\mu \nu}_{P_n}-i\frac{\eta^{\mu \nu}}{m_n^2}.
\end{align}
This restores gauge invariance in the otherwise nontransverse amplitude
\begin{align}
    \M^{\mu\nu \alpha \beta}|_{\F^T_3}=\text{tr}\Q^2 \sum_{m, n}c^{(3)}_{mn} F_n F_m \left(G_{P_n}^{\mu \alpha}(q_1)G_{P_m}^{\nu \beta}(q_2)+G_{P_n}^{\mu \beta}(q_1)G_{P_m}^{\nu \alpha}(q_2)\right)
\end{align}
and leads to a TFF $\F^T_3$ in the form \eqref{eq:FT3modesum} which is
regular at $q_i^2=0$.

However, in contrast to the antisymmetric tensor formulation of vector mesons,
this formulation misses the possibility to represent
interactions leading to a TFF $\F^T_1$ directly
in the form \eqref{eq:FT1modesum} by a local action.
Indeed, in the 4D version of the hQCD model, which is 
formulated in terms of Proca fields, \eqref{eq:FT1modesum}
arises from the local action \eqref{eq:STVVFT1} only due to a particular
interplay of the coupling constants $c_{mn}$, $c_{\gamma n}$
and $c_{\gamma\gamma}$ reflecting complete VMD in hQCD.

\section{$\hat{\Pi}_i$ functions for $\F^T_{1,3}$ with and without non-pole terms}

\label{appB}

Let us briefly recall that the general analysis outlined in \cite{Colangelo:2015ama, Colangelo:2017fiz} leads to the following  master formula for the HLbL contribution 
\begin{align}\label{master}
	a_\mu^\mathrm{HLbL}& = \frac{2 \alpha^3}{3 \pi^2} \int_0^\infty dQ_1 \int_0^\infty dQ_2 \int_{-1}^1 d\tau \sqrt{1-\tau^2} Q_1^3 Q_2^3\nonumber\\
	&\qquad\times\sum_{i=1}^{12} \bar T_i(Q_1,Q_2,\tau) \bar \Pi_i(Q_1,Q_2,Q_3)\,,
\end{align}
where $Q_1$ and $Q_2$ are the radial components of the Euclidean momenta. The hadronic scalar functions $\bar \Pi_i$ are  evaluated for the reduced kinematics
\begin{align}
(q_1^2,q_2^2,q_3^2,q_4^2)=(-Q_1^2,-Q_2^2,-Q_3^2=- Q_1^2 - 2 Q_1 Q_2 \tau - Q_2^2,0)\,.
\end{align}
(The complete list of the integral kernels $\bar T_i(Q_1,Q_2,\tau)$ can be found in Appendix B of~\cite{Colangelo:2017fiz}.)

The twelve scalar function $\bar \Pi_i$, could in principle be obtained for any HLbL tensor, {\it i.e.} the  correlation function of four electromagnetic quark currents, obtained in any model provided it respects  Lorentz and gauge invariance, crossing symmetries and some analyticity  properties. 

The extraction of the scalar functions $\bar \Pi_i$, is however a nontrivial task, requiring the decomposition of the HLbL tensor in particular basis of gauge invariant tensor structures.

This process produces certain ambiguities when one considers pole contributions to the HLbL tensor obtained by the exchange of single resonance, as is the case of the tensor resonances we are considering. 

Only recently, a new optimized basis has been introduced in ref.~\cite{Hoferichter:2024fsj} and explicit nonambiguous formula have been found for the exchange of spin-1 and spin-2 resonances, in terms of their TFF and propagators. Actually, for tensor particle exchange, the formulas hold when the tensor TFF has simplified structures. Beyond the simplest case in which the only nonvanishing coefficient is $\F_1$, as predicted by the Quark Model \cite{Schuler:1997yw}, unambiguous expressions in the dispersive framework for the $\bar \Pi_i$ can be written if either only $\F_{2,3}$ or $\F_{1,3}$ are nonvanishing,\footnote{In triangle kinematics, the problem this poses
for the dispersive approach can be avoided in the strategy developed in \cite{Ludtke:2023hvz,Ludtke:2024ase}.}
the latter  case precisely occurring in the hQCD model.
We stress however that these ambiguities do not play a role in a full evaluation including non-pole terms. In that case any HLbL tensor gives rise to unambiguous $\bar{\Pi}_i(Q_1,Q_2,Q_3)$. The data needed to define non-pole terms goes beyond the TFF where the meson is on-shell. As will be seen below, hQCD generates new Lorentz structures that are not present in the dispersive approach, since they do not contribute on-shell.

The functions $\bar\Pi_i$ can be assembled from six functions
$\hat\Pi_i$, $i=1,4,7,17,39,54$, evaluated below, cf.\ ref.\ \cite{Colangelo:2017fiz}.

\subsection{With non-pole terms}

The full amplitude including trace terms of
metric fluctuations reads
\begin{align}
\label{eq:fullTFF}
    \M^{\mu \nu}{}_{\alpha \beta}(q_1,q_2)= (T_1)^{\mu \nu }{}_{\alpha \beta} \left(-\frac{\text{tr} \Q^2}{g_5^2}\right)\int \frac{dz}{z} h(z) \J(z,q_1^2)\J(z,q_2) \nonumber\\
    + (T_3)^{\mu \nu }{}_{\alpha \beta} \left(-\frac{\text{tr} \Q^2}{g_5^2}\right)\int \frac{dz}{z} h(z) \frac{\partial_z \J(z,q_1^2)}{q_1^2}\frac{\partial_z \J(z,q_2^2)}{q_2^2} \nonumber\\
    + \eta_{\alpha \beta} P^{\mu}_{\;\sigma}(q_1)P^{\nu}_{\;\rho}(q_2)\eta^{\sigma \rho}\left(\frac{\text{tr} \Q^2}{g_5^2}\right)\int \frac{dz}{z} h(z) \partial_z \J(z,q_1^2)\partial_z \J(z,q_2^2) \nonumber\\
    +\frac{1}{2} \eta_{\alpha \beta} \left((q_1)_{\sigma}P^{\mu}_{\;\rho}(q_1)-(q_1)_{\rho}P^{\mu}_{\;\sigma}(q_1)\right)\left(q_2^{\sigma} P^{\nu \rho}(q_2)-q_2^{\rho} P^{\nu \sigma}(q_2)\right)  \nonumber\\
 \times\left(\frac{\text{tr} \Q^2}{g_5^2}\right)\int \frac{dz}{z} h(z)\J(z,q_1^2)\J(z,q_2^2) ,
\end{align}
which can be concisely summarised as
\begin{align}
     \M^{\mu \nu}_{\;\;\;\;\;\alpha \beta}(q_1,q_2)&= \left( (T^T_1)^{\mu \nu }_{\;\;\;\;\;\alpha \beta}-(T_1^S)^{\mu \nu}\eta_{\alpha \beta}\right)  F_1(q_1^2,q_2^2) \nonumber \\
    &+\left( (T^T_3)^{\mu \nu }_{\;\;\;\;\;\alpha \beta}-(T_2^S)^{\mu \nu}\eta_{\alpha \beta}\right)  F_3(q_1^2,q_2^2)
\end{align}
with $F_1= \frac{1}{m_T}\mathcal{F}_1, \, F_3= \frac{1}{m_T^3} \mathcal{F}_3$ and using the Lorentz structures $(T_i^S)^{\mu \nu}$ of \cite{Hoferichter:2020lap} used in the scalar TFF. They are given by
\begin{align}
    (T_1^S)^{\mu \nu}&= q_1 \cdot q_2 \eta^{\mu \nu}-q_2^{\mu}q_1^{\nu}, \\
     (T_2^S)^{\mu \nu}&= q_1^2 q_2^2 \eta^{\mu \nu} + q_1 \cdot q_2 q_1^{\mu} q_2^{\nu}- q_1^2 q_2^{\mu} q_2^{\nu}- q_2^2 q_1^{\mu} q_1^{\nu}.
\end{align}

The full hQCD HLbL tensor is  built up from two $\M$ vertices and one Fierz-Pauli propagator \eqref{eq:FP-prop} and reads
\begin{align}
    \Pi^{\mu_1 \mu_2 \mu_3 \mu_4}(q_1,q_2,q_3,q_4)=\, 
    &i\M^{\mu_1 \mu_2}_{\;\;\;\;\;\;\;\;\alpha \beta}(q_1,q_2) iG_T^{\alpha \beta \gamma \delta}(q_1+q_2)i\M^{\mu_3 \mu_4}_{\;\;\;\;\;\;\;\;\gamma \delta}(q_3,q_4) \nonumber \\
    +&i\M^{\mu_1 \mu_3}_{\;\;\;\;\;\;\;\;\alpha \beta}(q_1,q_3) iG_T^{\alpha \beta \gamma \delta}(q_1+q_3)i\M^{\mu_2 \mu_4}_{\;\;\;\;\;\;\;\;\gamma \delta}(q_2,q_4) \nonumber \\
    +&i\M^{\mu_1 \mu_4}_{\;\;\;\;\;\;\;\;\alpha \beta}(q_1,q_4) iG_T^{\alpha \beta \gamma \delta}(q_1+q_4)i\M^{\mu_2 \mu_3}_{\;\;\;\;\;\;\;\;\gamma \delta}(q_2,q_3).
\end{align}
The Fierz-Pauli propagator is only traceless in $\alpha \beta$ on-shell; off-shell the trace is given by a contact term, which makes the trace terms above contribute to the longitudinal part of  the $a_\mu$ contributions.
Using the projection techniques of \cite{Colangelo:2015ama,Colangelo:2017fiz,Hoferichter:2024fsj} one may calculate the relevant structure functions $\hat{\Pi}_i$ which can be straightforwardly used in the master formula for the $g-2$.
They read: 

\begin{align*}
\hat{\Pi}_1 = & \mathcal{F}_1(q_2^2, 0) \left( -\frac{2 \mathcal{F}_1(q_1^2, q_3^2)}{m_T^4} + \frac{2 q_1^2 \mathcal{F}_3(q_1^2, q_3^2)}{m_T^6} \right) \\
& + \mathcal{F}_1(q_1^2, 0) \left( -\frac{2 \mathcal{F}_1(q_2^2, q_3^2)}{m_T^4} + \frac{2 q_2^2 \mathcal{F}_3(q_2^2, q_3^2)}{m_T^6} \right),
\end{align*}

\begin{align*}
\hat{\Pi}_4 =  \frac{1}{3 m_T^8} \mathcal{F}_1(q_3^2, 0) & \biggl( 2 m_T^2 q_3^2 \mathcal{F}_1(q_1^2, q_2^2) - \Bigl( (q_1^2 - q_2^2)^2 + (q_1^2 + q_2^2) q_3^2  \\
& \quad +  m_T^2 \left( q_1^2 + q_2^2 + q_3^2 \right) \Bigr) \mathcal{F}_3(q_1^2, q_2^2) \biggr) \\
 + \frac{1}{3 m_T^8 (m_T^2 - q_3^2)} & 2 \mathcal{F}_1(q_3^2, 0) \biggl( 2 m_T^2 q_3^2 (m_T^2 + q_3^2) \mathcal{F}_1(q_1^2, q_2^2)  \\
& \quad + \Bigl( 2 m_T^4 (q_1^2 + q_2^2 + q_3^2) - q_3^2 \left( (q_1^2 - q_2^2)^2 + (q_1^2 + q_2^2) q_3^2 \right)  \\
& \quad - m_T^2 \left( (q_1^2 - q_2^2)^2 + 2 (q_1^2 + q_2^2) q_3^2 + q_3^4 \right) \Bigr) \mathcal{F}_3(q_1^2, q_2^2) \biggr) \\
 + \frac{1}{m_T^8 - m_T^6 q_2^2} & \mathcal{F}_1(q_2^2, 0) \biggl( m_T^2 \left( -2 m_T^2 + q_1^2 + q_2^2 + q_3^2 \right) \mathcal{F}_1(q_1^2, q_3^2)  \\
& \quad - \left( m_T^2 (q_1^2 - q_2^2 - q_3^2) + q_3^2 (q_1^2 + q_2^2 + q_3^2) \right) \mathcal{F}_3(q_1^2, q_3^2) \biggr) \\
 + \frac{1}{m_T^8 - m_T^6 q_1^2} &\mathcal{F}_1(q_1^2, 0) \biggl( m_T^2 \left( -2 m_T^2 + q_1^2 + q_2^2 + q_3^2 \right) \mathcal{F}_1(q_2^2, q_3^2)  \\
& \quad + \left( m_T^2 (q_1^2 - q_2^2 + q_3^2) - q_3^2 (q_1^2 + q_2^2 + q_3^2) \right) \mathcal{F}_3(q_2^2, q_3^2) \biggr),
\end{align*}

\begin{align*}
\hat{\Pi}_7 = & -\frac{2 (m_T^2 - q_1^2 + q_2^2) \mathcal{F}_1(q_3^2, 0) \mathcal{F}_3(q_1^2, q_2^2)}{m_T^8 - m_T^6 q_3^2} \\
& + \frac{\mathcal{F}_1(q_2^2, 0) \left( -2 m_T^2 \mathcal{F}_1(q_1^2, q_3^2) + 2 (m_T^2 + q_3^2) \mathcal{F}_3(q_1^2, q_3^2) \right)}{m_T^8 - m_T^6 q_2^2} \\
& - \frac{2 \mathcal{F}_1(q_1^2, 0) \mathcal{F}_3(q_2^2, q_3^2)}{m_T^6 - m_T^4 q_1^2},
\end{align*}

\begin{align*}
\hat{\Pi}_{17} = & -\frac{1}{3 m_T^8} \Bigg( \frac{2 m_T^2 (3 m_T^2 + 2 q_3^2) \mathcal{F}_1(q_1^2, q_2^2) \mathcal{F}_1(q_3^2, 0)}{m_T^2 - q_3^2} \\
& \quad - \frac{1}{m_T^2 - q_3^2} \left( 3 m_T^2 (q_1^2 + q_2^2) + 2 (m_T^2 + q_1^2 + q_2^2) q_3^2 \right) \mathcal{F}_1(q_3^2, 0) \mathcal{F}_3(q_1^2, q_2^2) \\
& \quad + 3 m_T^2 \left( \frac{(q_1^2 - q_3^2) \mathcal{F}_1(q_2^2, 0) \mathcal{F}_3(q_1^2, q_3^2)}{m_T^2 - q_2^2} + \frac{(q_2^2 - q_3^2) \mathcal{F}_1(q_1^2, 0) \mathcal{F}_3(q_2^2, q_3^2)}{m_T^2 - q_1^2} \right) \Bigg).
\end{align*}

\begin{align*}
\hat{\Pi}_{39} = & \frac{ \mathcal{F}_1(q_3^2, 0) \left( 2 m_T^2 \mathcal{F}_1(q_1^2, q_2^2) + \left( 2 m_T^2 - q_1^2 - q_2^2 \right) \mathcal{F}_3(q_1^2, q_2^2) \right)}{m_T^8 - m_T^6 q_3^2} \\
& + \frac{ \mathcal{F}_1(q_2^2, 0) \left( 2 m_T^2 \mathcal{F}_1(q_1^2, q_3^2) + \left( 2 m_T^2 - q_1^2 - q_3^2 \right) \mathcal{F}_3(q_1^2, q_3^2) \right)}{m_T^8 - m_T^6 q_2^2} \\
& + \frac{ \mathcal{F}_1(q_1^2, 0) \left( 2 m_T^2 \mathcal{F}_1(q_2^2, q_3^2) + \left( 2 m_T^2 - q_2^2 - q_3^2 \right) \mathcal{F}_3(q_2^2, q_3^2) \right)}{m_T^8 - m_T^6 q_1^2},
\end{align*}

\begin{align}
\hat{\Pi}_{54} = & \frac{ (q_1^2 - q_2^2) \mathcal{F}_1(q_3^2, 0) \mathcal{F}_3(q_1^2, q_2^2) }{ m_T^6 \left( m_T^2 - q_3^2 \right) } \nonumber\\
& + \frac{ \mathcal{F}_1(q_2^2, 0) \left( 2 m_T^2 \mathcal{F}_1(q_1^2, q_3^2) - q_1^2 \mathcal{F}_3(q_1^2, q_3^2) - q_3^2 \mathcal{F}_3(q_1^2, q_3^2) \right) }{ m_T^6 \left( m_T^2 - q_2^2 \right) } \nonumber\\
& - \frac{ \mathcal{F}_1(q_1^2, 0) \left( 2 m_T^2 \mathcal{F}_1(q_2^2, q_3^2) - q_2^2 \mathcal{F}_3(q_2^2, q_3^2) - q_3^2 \mathcal{F}_3(q_2^2, q_3^2) \right) }{ m_T^6 \left( m_T^2 - q_1^2 \right) }.
\end{align}

\subsection{Pole terms only}

For the sake of comparison, we also include below the
pole-term parts as defined by the dispersive procedure in the optimized basis of ref.~\cite{Hoferichter:2024fsj}.

\begin{align*}
\hat{\Pi}_1=\mathcal{F}_1(q_2^2, 0) \left( - \frac{\mathcal{F}_1(q_1^2, q_3^2)}{m_T^4} + \frac{(-m_T^2 + q_1^2) \mathcal{F}_3(q_1^2, q_3^2)}{m_T^6} \right) &+ \\
\mathcal{F}_1(q_1^2, 0) \left( - \frac{\mathcal{F}_1(q_2^2, q_3^2)}{m_T^4} + \frac{(-m_T^2 + q_2^2) \mathcal{F}_3(q_2^2, q_3^2)}{m_T^6} \right),
\end{align*}

\begin{align*}
\hat{\Pi}_4 =  \frac{1}{3 m_T^8 (m_T^2 - q_3^2)} &
\mathcal{F}_1(q_3^2, 0) \Bigg( 
    8 m_T^4 q_3^2 \mathcal{F}_1(q_1^2, q_2^2) + \\
    &\left( 6 m_T^6 - 3 m_T^2 (q_1^2 - q_2^2)^2 + 3 m_T^4 (q_1^2 + q_2^2) - \right. \\
    &\left( 4 m_T^4 + (q_1^2 - q_2^2)^2 + 5 m_T^2 (q_1^2 + q_2^2) \right) q_3^2 ) \mathcal{F}_3(q_1^2, q_2^2) \Bigg) + \\
\frac{1}{m_T^8 - m_T^6 q_2^2} & \mathcal{F}_1(q_2^2, 0) \Bigg( 
    m_T^2 (-4 m_T^2 + q_1^2 + 3 q_2^2 + q_3^2) \mathcal{F}_1(q_1^2, q_3^2) + \\
    &\left( m_T^4 - q_1^2 q_2^2 + 2 m_T^2 q_3^2 - 
    q_3^2 (q_1^2 + 2 q_2^2 + q_3^2) \right) \mathcal{F}_3(q_1^2, q_3^2) \Bigg) + \\
\frac{1}{m_T^8 - m_T^6 q_1^2} & \mathcal{F}_1(q_1^2, 0) \Bigg( 
    m_T^2 (-4 m_T^2 + 3 q_1^2 + q_2^2 + q_3^2) \mathcal{F}_1(q_2^2, q_3^2) + \\
    &\left( m_T^4 - q_1^2 q_2^2 + 2 m_T^2 q_3^2 - 
    q_3^2 (2 q_1^2 + q_2^2 + q_3^2) \right) \mathcal{F}_3(q_2^2, q_3^2) \Bigg),
\end{align*}

\begin{align*}
\hat{\Pi}_7 = & - \frac{2 (m_T^2 - q_1^2 + q_2^2) \mathcal{F}_1(q_3^2, 0) \mathcal{F}_3(q_1^2, q_2^2)}{m_T^8 - m_T^6 q_3^2} + \\
& \frac{\mathcal{F}_1(q_2^2, 0) \left( -2 m_T^2 \mathcal{F}_1(q_1^2, q_3^2) + 2 (m_T^2 + q_3^2) \mathcal{F}_3(q_1^2, q_3^2) \right)}{m_T^8 - m_T^6 q_2^2} - \\
& \frac{\mathcal{F}_1(q_1^2, 0) \left( 2 m_T^2 \mathcal{F}_3(q_2^2, q_3^2) \right)}{m_T^8 - m_T^6 q_1^2},
\end{align*}

\begin{align*}
\hat{\Pi}_{17} = & \frac{1}{3 m_T^8 (m_T^2 - q_3^2)} \mathcal{F}_1(q_3^2, 0) \Big( -2 (3 m_T^4 + 2 m_T^2 q_3^2) \mathcal{F}_1(q_1^2, q_2^2) \\
& \quad + \left( 3 m_T^2 (q_1^2 + q_2^2) + 2 (m_T^2 + q_1^2 + q_2^2) q_3^2 \right) \mathcal{F}_3(q_1^2, q_2^2) \Big) \\
& + \frac{(-q_1^2 + q_3^2) \mathcal{F}_1(q_2^2, 0) \mathcal{F}_3(q_1^2, q_3^2)}{m_T^8 - m_T^6 q_2^2} \\
& + \frac{(-q_2^2 + q_3^2) \mathcal{F}_1(q_1^2, 0) \mathcal{F}_3(q_2^2, q_3^2)}{m_T^8 - m_T^6 q_1^2},
\end{align*}

\begin{align*}
\hat{\Pi}_{39} = & \frac{\mathcal{F}_1(q_3^2, 0) \left( 4 m_T^2 \mathcal{F}_1(q_1^2, q_2^2) + \left( 4 m_T^2 - 2 (q_1^2 + q_2^2) \right) \mathcal{F}_3(q_1^2, q_2^2) \right)}{2 m_T^6 (m_T^2 - q_3^2)} \\
& + \frac{\mathcal{F}_1(q_2^2, 0) \left( 4 m_T^2 \mathcal{F}_1(q_1^2, q_3^2) + \left( 4 m_T^2 - 2 (q_1^2 + q_3^2) \right) \mathcal{F}_3(q_1^2, q_3^2) \right)}{2 m_T^6 (m_T^2 - q_2^2)} \\
& + \frac{\mathcal{F}_1(q_1^2, 0) \left( 4 m_T^2 \mathcal{F}_1(q_2^2, q_3^2) + \left( 4 m_T^2 - 2 (q_2^2 + q_3^2) \right) \mathcal{F}_3(q_2^2, q_3^2) \right)}{2 m_T^6 (m_T^2 - q_1^2)},
\end{align*}

\begin{align}
\hat{\Pi}_{54} = & \frac{(q_1^2 - q_2^2) \mathcal{F}_1(q_3^2, 0) \mathcal{F}_3(q_1^2, q_2^2)}{m_T^8 - m_T^6 q_3^2} \nonumber\\
& + \frac{\mathcal{F}_1(q_2^2, 0) \left( 2 m_T^2 \mathcal{F}_1(q_1^2, q_3^2) - (q_1^2 + q_3^2) \mathcal{F}_3(q_1^2, q_3^2) \right)}{m_T^8 - m_T^6 q_2^2} \nonumber\\
& + \frac{\mathcal{F}_1(q_1^2, 0) \left( -2 m_T^2 \mathcal{F}_1(q_2^2, q_3^2) + (q_2^2 + q_3^2) \mathcal{F}_3(q_2^2, q_3^2) \right)}{m_T^8 - m_T^6 q_1^2}.
\end{align}

\section{$\hat{\Pi}_i$ functions with tensor bulk-to-bulk propagator}
\label{appC}
\newcommand\Jpq{\widetilde{\mathcal{J}}}

With $\Jpq(z,q^2)={\J'(z,q^2)}/{q^2}$ 
and $G(q^2,z,z')$ the tensor bulk-to-bulk propagator
with the combinations
\begin{align}
    G_1(q^2,z,z')
    \equiv& \frac{1}{q^2}\left( G(q^2,z,z')-G(0,z,z') \right),\\
     G_2(q^2,z,z')
     \equiv&\frac{1}{q^2}\left( G_1(q^2,z,z')-G_1(0,z,z') \right),
\end{align}
the sum over the infinite tower of (full) tensor meson contributions yields
\begin{align}\label{Pi1TGreen}
\hat\Pi_1 =&\frac{1}{k_T}\left(\frac{\text{tr}(\Q^2)}{g_5^2}\right)^2 \iint_0^{z_0}dz_1\,dz_2 
\frac{2}{z_1 z_2}  G(0, z_1, z_2)  
\nonumber\\&\Bigg(\Big( 
        \mathcal{J}(z_1, q_2^2) \mathcal{J}(z_2, q_1^2)  
        + \mathcal{J}(z_1, q_1^2) \mathcal{J}(z_2, q_2^2) 
    \Big) 
    \mathcal{J}(z_2, q_3^2) \nonumber \\
& - 
    \Big( 
        q_1^2 \mathcal{J}(z_1, q_2^2) \Jpq(z_2, q_1^2)  
        + q_2^2 \mathcal{J}(z_1, q_1^2) \Jpq(z_2, q_2^2) 
    \Big) 
    \Jpq(z_2, q_3^2)  
\Bigg),
\end{align}

\begin{align}
   \hat\Pi_4 =& \frac{1}{k_T}\left(\frac{\text{tr}(\Q^2)}{g_5^2}\right)^2\iint_0^{z_0}dz_1\,dz_2  \frac{1}{3 {z_1} {z_2}}\nonumber\\
   &\Bigg(-2 \J({z_1},{q_3^2})
   \Big({\Jpq}({z_2},{q_1^2}) {\Jpq}({z_2},{q_2^2})
   \Big(2 ({q_1^2}+{q_2^2}+{q_3^2})
   G({q_3^2},{z_1},{z_2})\nonumber\\&-\Big(2 {q_3^2}
   ({q_1^2}+{q_2^2})+({q_1^2}-{q_2^2})^2+{q_3^2}^2\Big)
   {G_1}({q_3^2},{z_1},{z_2})\nonumber\\&-{q_3^2} \Big({q_3^2}
   ({q_1^2}+{q_2^2})+({q_1^2}-{q_2^2})^2\Big)
   {G_2}({q_3^2},{z_1},{z_2})\Big)\nonumber\\&+2 {q_3^2}
   \J({z_2},{q_1^2}) \J({z_2},{q_2^2})
   \big({G_1}({q_3^2},{z_1},{z_2})+{q_3^2}
   {G_2}({q_3^2},{z_1},{z_2})\big)\Big)\nonumber\\&+\J({z_1},{q_3^2}
   ) \Big({\Jpq}({z_2},{q_1^2})
   {\Jpq}({z_2},{q_2^2})
   \Big(({q_1^2}+{q_2^2}+{q_3^2})
   G(0,{z_1},{z_2})\nonumber\\&+\Big({q_3^2}
   ({q_1^2}+{q_2^2})+({q_1^2}-{q_2^2})^2\Big)
   {G_1}(0,{z_1},{z_2})\Big)-2 {q_3^2}
   {G_1}(0,{z_1},{z_2}) \J({z_2},{q_1^2})
   \J({z_2},{q_2^2})\Big)\nonumber\\&+3 \J({z_1},{q_2^2})
   \bigg(G({q_2^2},{z_1},{z_2}) \Big(({q_1^2}-{q_2^2}-{q_3^2})
   {\Jpq}({z_2},{q_1^2}) {\Jpq}({z_2},{q_3^2})+2
   \J({z_2},{q_1^2})
   \J({z_2},{q_3^2})\Big)\nonumber\\&+({q_1^2}+{q_2^2}+{q_3^2})
   {G_1}({q_2^2},{z_1},{z_2}) \Big({q_3^2}
   {\Jpq}({z_2},{q_1^2})
   {\Jpq}({z_2},{q_3^2})-\J({z_2},{q_1^2})
   \J({z_2},{q_3^2})\Big)\bigg)\nonumber\\&+3 \J({z_1},{q_1^2})
   \bigg(G({q_1^2},{z_1},{z_2}) \Big(2 \J({z_2},{q_2^2})
   \J({z_2},{q_3^2})-({q_1^2}-{q_2^2}+{q_3^2})
   {\Jpq}({z_2},{q_2^2})
   {\Jpq}({z_2},{q_3^2})\Big)\nonumber\\&+({q_1^2}+{q_2^2}+{q_3^2})
   {G_1}({q_1^2},{z_1},{z_2}) \big({q_3^2}
   {\Jpq}({z_2},{q_2^2})
   {\Jpq}({z_2},{q_3^2})-\J({z_2},{q_2^2})
   \J({z_2},{q_3^2})\big)\bigg)\Bigg), \nonumber
\end{align}

\begin{align}
  \hat\Pi_7 =& \frac{1}{k_T}\left(\frac{\text{tr}(\Q^2)}{g_5^2}\right)^2\iint_0^{z_0}dz_1\,dz_2   \frac{2}{{z_1} {z_2}} 
  \nonumber\\&\Bigg({\Jpq}({z_2},{q_1^2})
   {\Jpq}({z_2},{q_2^2}) \J({z_1},{q_3^2})
   \Big(G({q_3^2},{z_1},{z_2}) 
   +({q_2^2}-{q_1^2})
   {G_1}({q_3^2},{z_1},{z_2})\Big)\nonumber\\&-{\Jpq}({z_2},{q_1^2}) \J({z_1},{q_2^2}) {\Jpq}({z_2},{q_3^2})
   \Big(G({q_2^2},{z_1},{z_2})+{q_3^2}
   {G_1}({q_2^2},{z_1},{z_2})\Big)\nonumber\\&+\J({z_1},{q_1^2})
   {\Jpq}({z_2},{q_2^2}) {\Jpq}({z_2},{q_3^2})
   G({q_1^2},{z_1},{z_2})\nonumber\\&+\J({z_2},{q_1^2})
   \J({z_1},{q_2^2}) \J({z_2},{q_3^2})
   {G_1}({q_2^2},{z_1},{z_2})\Bigg), \nonumber
\end{align}

\begin{align}
  \hat\Pi_{17} =&\frac{1}{k_T}\left(\frac{\text{tr}(\Q^2)}{g_5^2}\right)^2\iint_0^{z_0}dz_1\,dz_2  \frac{1}{3 {z_1} {z_2}} \nonumber\\&\Bigg({G_1}({q_3^2},{z_1},{z_2}) \Big(6 \J({z_2},0) \J({z_1},{q_1^2}) \J({z_1},{q_2^2}) \J({z_2},{q_3^2})\nonumber\\&- (3
   ({q_1^2}+{q_2^2})+2 {q_3^2}) {\Jpq}({z_2},{q_1^2}) {\Jpq}({z_2},{q_2^2}) \J({z_1},{q_3^2})\Big) \nonumber\\&+ 3
   {\Jpq}({z_2},{q_3^2}) \Big(({q_1^2}-{q_3^2}) {\Jpq}({z_2},{q_1^2}) \J({z_1},{q_2^2})
   {G_1}({q_2^2},{z_1},{z_2})\nonumber \\&+({q_2^2}-{q_3^2}) \J({z_1},{q_1^2}) {\Jpq}({z_2},{q_2^2}) {G_1}({q_1^2},{z_1},{z_2})\Big)\nonumber\\&+2
   {q_3^2} {G_2}({q_3^2},{z_1},{z_2}) \Big(2 \J({z_2},0) \J({z_1},{q_1^2}) \J({z_1},{q_2^2}) \J({z_2},{q_3^2})\nonumber\\&-({q_1^2}+{q_2^2})
    {\Jpq}({z_2},{q_1^2}) {\Jpq}({z_2},{q_2^2}) \J({z_1},{q_3^2})\Big) \Bigg) ,\nonumber
\end{align}

\begin{align}
   \hat\Pi_{39}=& \frac{1}{k_T}\left(\frac{\text{tr}(\Q^2)}{g_5^2}\right)^2\iint_0^{z_0}dz_1\,dz_2 \frac{1}{{z_1} {z_2}} 
   \nonumber\\&\Bigg(-2 \Big({\Jpq}({z_2},{q_1^2}) {\Jpq}({z_2},{q_2^2}) \J({z_1},{q_3^2})
   G({q_3^2},{z_1},{z_2})\nonumber \\&+\J({z_2},{q_3^2}) \big(\J({z_2},{q_1^2}) \J({z_1},{q_2^2})
   {G_1}({q_2^2},{z_1},{z_2})\nonumber \\&+\J({z_1},{q_1^2}) \J({z_2},{q_2^2}) {G_1}({q_1^2},{z_1},{z_2})\big)\Big)\nonumber \\&+{\Jpq}({z_2},{q_3^2})
   \Big({\Jpq}({z_2},{q_1^2}) \J({z_1},{q_2^2}) (({q_1^2}+{q_3^2}) {G_1}({q_2^2},{z_1},{z_2})\nonumber \\&-2
   G({q_2^2},{z_1},{z_2}))+\J({z_1},{q_1^2}) {\Jpq}({z_2},{q_2^2}) (({q_2^2}+{q_3^2}) {G_1}({q_1^2},{z_1},{z_2})-2
   G({q_1^2},{z_1},{z_2}))\Big)\nonumber \\&+\J({z_1},{q_3^2}) {G_1}({q_3^2},{z_1},{z_2}) \left(({q_1^2}+{q_2^2}) {\Jpq}({z_2},{q_1^2})
   {\Jpq}({z_2},{q_2^2})-2 \J({z_2},{q_1^2}) \J({z_2},{q_2^2})\right)\Bigg),\nonumber
\end{align}

\begin{align}
  \hat\Pi_{54} =& \frac{1}{k_T}\left(\frac{\text{tr}(\Q^2)}{g_5^2}\right)^2\iint_0^{z_0}dz_1\,dz_2   \frac{1}{{z_1} {z_2}} \nonumber\\&\Bigg(({q_2^2}-{q_1^2}) {\Jpq}({z_2},{q_1^2}) {\Jpq}({z_2},{q_2^2}) \J({z_1},{q_3^2})
   {G_1}({q_3^2},{z_1},{z_2})\nonumber\\&+\J({z_1},{q_2^2}) {G_1}({q_2^2},{z_1},{z_2}) \Big(({q_1^2}+{q_3^2}) {\Jpq}({z_2},{q_1^2})
   {\Jpq}({z_2},{q_3^2})\nonumber\\&-2 \J({z_2},{q_1^2}) \J({z_2},{q_3^2})\Big)+\J({z_1},{q_1^2}) {G_1}({q_1^2},{z_1},{z_2}) \Big(2
   \J({z_2},{q_2^2}) \J({z_2},{q_3^2})\nonumber\\&-({q_2^2}+{q_3^2}) {\Jpq}({z_2},{q_2^2}) {\Jpq}({z_2},{q_3^2})\Big)\Bigg) . 
\end{align}

\bibliographystyle{JHEP}
\bibliography{SM}

\end{document}